\documentclass{article}
\usepackage{amsmath,amssymb,bm}

\usepackage{graphicx}        
\usepackage[bottom]{footmisc}

\title{\bf \Large Binary Systems as Test-beds of Gravity Theories\footnote{
Based on lectures given at the SIGRAV School ``A Century from Einstein Relativity:
Probing Gravity Theories in Binary Systems'', Villa Olmo (Como Lake, Italy),
17-21 May 2005. To appear in the Proceedings, edited by M. Colpi et al.
(to be published by Springer).}}
\author{Thibault Damour \\ { \ } \\
Institut des Hautes Etudes Scientifiques, 35 route de Chartres, \\ F-91440 Bures-sur-Yvette, France
}
\date{ \ }
\begin{document}
\maketitle

We review the general relativistic theory of the motion, and of the timing, of binary systems containing compact objects (neutron stars or black holes). Then we indicate the various ways one can use binary pulsar data to test the strong-field and/or radiative aspects of General Relativity, and of general classes of alternative theories of relativistic gravity.

\section{Introduction}
\label{sec:1}

The discovery of binary pulsars in 1974 \cite{HulseTaylor} opened up a new testing ground for relativistic gravity. Before this discovery, the  only available testing ground for relativistic gravity was the solar system. As Einstein's theory of General Relativity (GR) is one of the basic pillars of modern science, it deserves to be tested, with the highest possible accuracy, in all its aspects. In the solar system, the gravitational field is slowly varying and represents only a very small deformation of a flat spacetime. As a consequence, solar system tests can only probe the quasi-stationary (non radiative) {\it weak-field} limit of relativistic gravity. By contrast binary systems containing compact objects (neutron stars or black holes) involve spacetime domains (inside and near the compact objects) where {\it the gravitational field is strong}. Indeed, the surface relativistic gravitational field $h_{00} \simeq 2 \, GM/c^2 R$ of a neutron star is of order $0.4$, which is close to the one of a black hole ($2 \, GM/c^2 R = 1$) and much larger than the surface gravitational fields of solar system bodies: $(2 \, GM/c^2 R)_{\rm Sun} \sim 10^{-6}$, $(2 \, GM/c^2 R)_{\rm Earth} \sim 10^{-9}$. In addition, the high stability of ``pulsar clocks'' has made it possible to monitor the dynamics of its orbital motion down to a precision allowing one to measure the small $(\sim (v/c)^5)$ orbital effects linked to the propagation of the gravitational field at the velocity of light between the pulsar and its companion.

\smallskip

The recent discovery of the remarkable {\it double} binary pulsar PSR~J0737$-$ 3039 \cite{Burgay03,Lyne04} (see also the contributions of M.~Kramer and A.~Possenti to these proceedings) has renewed the interest in the use of binary pulsars as test-beds of gravity theories. The aim of these notes is to provide an introduction to the theoretical frameworks needed for interpreting binary pulsar data as tests of GR and alternative gravity theories.

\section{Motion of binary pulsars in general relativity}\label{sec:2}

The traditional (text book) approach to the problem of motion of $N$ separate bodies in GR consists of solving, by successive approximations, Einstein's field equations (we use the signature $-+++$)
\begin{equation}
\label{eq2.1}
R_{\mu\nu} - \frac{1}{2} \, R \, g_{\mu\nu} = \frac{8\pi \, G}{c^4} \, T_{\mu\nu} \, ,
\end{equation}
together with their consequence
\begin{equation}
\label{eq2.2}
\nabla_{\nu} \, T^{\mu\nu} = 0 \, .
\end{equation}
To do so, one assumes some specific matter model, say a perfect fluid,
\begin{equation}
\label{eq2.3}
T^{\mu\nu} = (\varepsilon + p) \, u^{\mu} \, u^{\nu} + p \, g^{\mu\nu} \, .
\end{equation}
One expands (say in powers of Newton's constant)
\begin{equation}
\label{eq2.4}
g_{\mu\nu} (x^{\lambda}) = \eta_{\mu\nu} + h_{\mu\nu}^{(1)} + h_{\mu\nu}^{(2)} + \ldots \, ,
\end{equation}
together with the use of the simplifications brought by the `Post-Newtonian' approximation ($\partial_0 \, h_{\mu\nu} = c^{-1} \, \partial_t \, h_{\mu\nu} \ll \partial_i \, h_{\mu\nu}$; $v/c \ll 1$, $p \ll \varepsilon$). Then one integrates the local material equation of motion (\ref{eq2.2}) over the volume of each separate body, labelled say by $a=1,2,\ldots , N$. In so doing, one must define some `center of mass' $z_a^i$ of body $a$, as well as some (approximately conserved) `mass' $m_a$ of body $a$, together with some corresponding `spin vector' $S_a^i$ and, possibly, higher multipole moments.

\smallskip

An important feature of this traditional method is to use a {\it unique coordinate chart} $x^{\mu}$ to describe the full $N$-body system. For instance, the center of mass, shape and spin of each body $a$ are all described within this common coordinate system $x^{\mu}$. This use of a single chart has several inconvenient aspects, even in the case of weakly self-gravitating bodies (as in the solar system case). Indeed, it means for instance that a body which is, say, spherically symmetric in its own `rest frame' $X^{\alpha}$ will appear as deformed into some kind of ellipsoid in the common coordinate chart $x^{\mu}$. Moreover, it is not clear how to construct `good definitions' of the center of mass, spin vector, and higher multipole moments of body $a$, when described in the common coordinate chart $x^{\mu}$. In addition, as we are interested in the motion of strongly self-gravitating bodies, it is not a priori justified to use a simple expansion of the type (\ref{eq2.4}) because $h_{\mu\nu}^{(1)} \sim \underset{a}{\sum} \, Gm_a / (c^2 \, \vert \bm{x} - \bm{z}_a \vert)$ will not be uniformly small in the common coordinate system $x^{\mu}$. It will be small if one stays far away from each object $a$, but, as recalled above, it will become of order unity on the surface of a compact body.

\smallskip

These two shortcomings of the traditional `one-chart' approach to the relativistic problem of motion can be cured by using a `{\it multi-chart' approach}.The multi-chart approach describes the motion of $N$ (possibly, but not necessarily, compact) bodies by using $N+1$ separate coordinate systems: (i) one {\it global} coordinate chart $x^{\mu}$ ($\mu = 0,1,2,3$) used to describe the spacetime outside $N$ `tubes', each containing one body, and (ii) $N$ {\it local} coordinate charts $X_a^{\alpha}$ ($\alpha = 0,1,2,3$; $a = 1,2,\ldots , N$) used to describe the spacetime in and around each body $a$. The multi-chart approach was first used to discuss the motion of black holes and other compact objects 
\cite{Manasse,DEath,Kates,Eardley75,WillEardley77,Damour83,Thorne-Hartle,Will93}. Then it was also found to be very convenient for describing, with the high-accuracy required for dealing with modern technologies such as VLBI, systems of $N$ weakly self-gravitating bodies, such as the solar system \cite{Brumberg-Kopeikin,DSX}.

\smallskip

The essential idea of the multi-chart approach is to combine the information contained in {\it several expansions}. One uses both a global expansion of the type (\ref{eq2.4}) and several local expansions of the type
\begin{equation}
\label{eq2.5}
G_{\alpha\beta} (X_a^{\gamma}) = G_{\alpha\beta}^{(0)} (X_a^{\gamma} ; m_a) + H_{\alpha\beta}^{(1)} (X_a^{\gamma} ; m_a , m_b) + \cdots \, ,
\end{equation}
where $G_{\alpha\beta}^{(0)} (X ; m_a)$ denotes the (possibly strong-field) metric generated by an isolated body of mass $m_a$ (possibly with the additional effect of spin).

\smallskip

The separate expansions (\ref{eq2.4}) and (\ref{eq2.5}) are then `matched' in some overlapping domain of common validity of the type $Gm_a / c^2 \lesssim R_a \ll \vert \bm{x} - \bm{z}_a \vert \ll d \sim \vert \bm{x}_a - \bm{x}_b \vert$ (with $b \ne a$), where one can relate the different coordinate systems by expansions of the form
\begin{equation}
\label{eq2.6}
x^{\mu} = z_a^{\mu} (T_a) + e_i^{\mu} (T_a) \, X_a^i + \frac{1}{2} \, f_{ij}^{\mu} (T_a) \, X_a^i \, X_a^j + \cdots
\end{equation}

The multi-chart approach becomes simplified if one considers {\it compact} bodies (of radius $R_a$ comparable to $2 \, Gm_a / c^2$). In this case, it was shown \cite{Damour83}, by considering how the `internal expansion' (\ref{eq2.5}) propagates into the `external' one (\ref{eq2.4}) via the matching (\ref{eq2.6}), that, {\it in General Relativity}, the internal structure of each compact body was {\it effaced} to a very high degree, when seen in the external expansion (\ref{eq2.4}). For instance, for non spinning bodies, the internal structure of each body (notably the way it responds to an external tidal excitation) shows up in the external problem of motion only at the {\it fifth post-Newtonian} (5PN) approximation, i.e. in terms of order $(v/c)^{10}$ in the equations of motion.

\smallskip

This {\it `effacement of internal structure'} indicates that it should be possible to simplify the rigorous multi-chart approach by skeletonizing each compact body by means of some delta-function source. Mathematically, the use of distributional sources is delicate in a nonlinear theory such as GR. However, it was found that one can reproduce the results of the more rigorous matched-multi-chart approach by treating the divergent integrals generated by the use of delta-function sources by means of (complex) analytic continuation \cite{Damour83}. The most efficient method (especially to high PN orders) has been found to use analytic continuation in the dimension of space $d$ \cite{tHooftVeltman}.

\smallskip

Finally, the most efficient way to derive the general relativistic equations of motion of $N$ compact bodies consists of solving the equations derived from the action (where $g \equiv -\det (g_{\mu\nu})$)
\begin{equation}
\label{eq2.7}
S = \int \frac{d^{d+1} \, x}{c} \ \sqrt g \ \frac{c^4}{16\pi \, G} \ R(g) - \sum_a m_a \, c \int \sqrt{-g_{\mu\nu} (z_a^{\lambda}) \, dz_a^{\mu} \, dz_a^{\nu}} \, ,
\end{equation}
formally using the standard weak-field expansion (\ref{eq2.4}), but considering the space dimension $d$ as an arbitrary complex number which is sent to its physical value $d=3$ only at the end of the calculation.

\smallskip

Using this method\footnote{Or, more precisely, an essentially equivalent analytic continuation using the so-called `Riesz kernels'.} one has derived the equations of motion of {\it two} compact bodies at the 2.5PN $(v^5 / c^5)$ approximation level needed for describing binary pulsars \cite{DamourDeruelle81,Damour82,Damour83}:
\begin{eqnarray}
\label{eq2.8}
\frac{d^2 \, z_a^i}{dt^2} &= &A_{a0}^i (\bm{z}_a - \bm{z}_b) + c^{-2} \, A_{a2}^i (\bm{z}_a - \bm{z}_b , \bm{v}_a , \bm{v}_b) \nonumber \\
&+ &c^{-4} \, A_{a4}^i (\bm{z}_a - \bm{z}_b , \bm{v}_a , \bm{v}_b , \bm{S}_a , \bm{S}_b) \nonumber \\
&+ &c^{-5} \, A_{a5}^i (\bm{z}_a - \bm{z}_b , \bm{v}_a - \bm{v}_b) + {\mathcal O} (c^{-6}) \, .
\end{eqnarray}
Here $A_{a0}^i = - Gm_b (z_a^i - z_b^i) / \vert z_a - z_b \vert^3$ denotes the Newtonian acceleration, $A_{a2}^i$ its 1PN modification, $A_{a4}^i$ its 2PN modification (together with the spin-orbit effects), and $A_{a5}^i$ the $2.5$PN contribution of order $v^5/c^5$. [See the references above; or the review \cite{Damour87}, for more references and the explicit expressions of $A_2$, $A_4$ and $A_5$.] It was verified that the term $A_{a5}^i$ has  the effect of decreasing the mechanical energy of the system by an amount equal (on average) to the energy lost in the form of gravitational wave flux at infinity. Note, however, that here $A_{a5}^i$ was derived, in the near zone of the system, as a {\it direct} consequence of the general relativistic propagation of gravity, at the velocity $c$, between the two bodies. This highlights the fact that binary pulsar tests of the existence of $A_{a5}^i$ are {\it direct tests of the reality of gravitational radiation}.

\smallskip

Recently, the equations of motion (\ref{eq2.8}) have been computed to even higher accuracy: 3PN $\sim v^6 / c^6$ \cite{JaranowskiSchafer98,BlanchetFaye01,DamourJaranowskiSchafer01,ItohFutamase03,BlanchetDamourEsposito-Farese04} and $3.5 {\rm PN} \sim v^7 / c^7$ \cite{PatiWill01,KonigsdorfferFayeSchafer03,NissankeBlanchet05} (see also the review \cite{BlanchetLR}). These refinements are, however, not (yet) needed for interpreting binary pulsar data.

\section{Timing of binary pulsars in general relativity}\label{sec:3}

In order to extract observational effects from the equations of motion (\ref{eq2.8}) one needs to go through two steps: (i) to solve the equations of motion (\ref{eq2.8}) so as to get the coordinate positions $\bm{z}_1$ and $\bm{z}_2$ as explicit functions of the coordinate time $t$, and (ii) to relate the coordinate motion $\bm{z}_a (t)$ to the pulsar observables, i.e. mainly to the times of arrival of electromagnetic pulses on Earth.

\smallskip

The first step has been accomplished, in a form particularly useful for discussing pulsar timing, in Ref.~\cite{DD85}. There (see also \cite{DamourPRL83}) it was shown that, when considering the full (periodic and secular) effects of the $A_2 \sim v^2 / c^2$ terms in Eq.~(\ref{eq2.8}), together with the secular effects of the $A_4 \sim v^4 / c^4$ and $A_5 \sim v^5 / c^5$ terms, the relativistic two-body motion could be written in a very simple `quasi-Keplerian' form (in polar coordinates), namely:
\begin{equation}
\label{eq3.1}
\int n \, dt + \sigma = u - e_t \sin u \, ,
\end{equation}
\begin{equation}
\label{eq3.2}
\theta - \theta_0 = (1+k) \, 2 \arctan \left[ \left( \frac{1+e_{\theta}}{1-e_{\theta}} \right)^{\frac{1}{2}} \tan \frac{u}{2} \right] \, ,
\end{equation}
\begin{eqnarray}
\label{eq3.3}
R &\equiv &r_{ab} = a_R (1-e_R \cos u) \, , \\
\label{eq3.4}
r_a &\equiv &\vert \bm{z}_a - \bm{z}_{CM} \vert = a_r (1-e_r \cos u) \, , \\
\label{eq3.5}
r_b &\equiv &\vert \bm{z}_b - \bm{z}_{CM} \vert = a_{r'} (1-e_{r'} \cos u) \, .
\end{eqnarray}

Here $n \equiv 2\pi / P_b$ denotes the orbital frequency, $k = \Delta\theta / 2\pi = \langle \dot\omega \rangle / n = \langle \dot\omega \rangle \, P_b / 2\pi$ the fractional periastron advance per orbit, $u$ an auxiliary angle (`relativistic eccentric anomaly'), $e_t , e_{\theta} , e_R , e_r$ and $e_{r'}$ various `relativistic eccentricities' and $a_R , a_r$ and $a_{r'}$ some `relativistic semi-major axes'. See \cite{DD85} for the relations between these quantities, as well as their link to the relativistic energy and angular momentum $E,J$. A direct study \cite{DamourPRL83} of the dynamical effect of the contribution $A_5 \sim v^5 / c^5$ in the equations of motion (\ref{eq2.8}) has checked that it led to a secular increase of the orbital frequency $n(t) \simeq n(0) + \dot n (t-t_0)$, and thereby to a quadratic term in the `relativistic mean anomaly' $\ell = \int n \, dt + \sigma$ appearing on the left-hand side (L.H.S.) of Eq.~(\ref{eq3.1}):
\begin{equation}
\label{eq3.6}
\ell \simeq \sigma_0 + n_0 (t-t_0) + \frac{1}{2} \, \dot n (t-t_0)^2 \, .
\end{equation}

As for the contribution $A_4 \sim v^4 / c^4$ it induces several secular effects in the orbital motion: various 2PN contributions to the dimensionless periastron parameter $k$ ($\delta_4 \, k \sim v^4 / c^4 +$ spin-orbit effects), and secular variations in the inclination of the orbital plane (due to spin-orbit effects).

\smallskip

The second step in relating (\ref{eq2.8}) to pulsar observations has been accomplished through the derivation of a {\it `relativistic timing formula'} \cite{BT76,DD86}. The `timing formula' of a binary pulsar is a multi-parameter mathematical function relating the observed time of arrival (at the radio-telescope) of the center of the $N$$^{\rm th}$ pulse to the integer $N$. It involves many different physical effects: (i) dispersion effects, (ii) travel time across the solar system, (iii) gravitational delay due to the Sun and the planets, (iv) time dilation effects between the time measured on the Earth and the solar-system-barycenter time, (v) variations in the travel time between the binary pulsar and the solar-system barycenter (due to relative accelerations, parallax and proper motion), (vi) time delays happening within the binary system. We shall focus here on the time delays which take place within the binary system (see the lectures of M.~Kramer for a discussion of the other effects).

\smallskip

For a proper derivation of the time delays occurring within the binary system we need to use the multi-chart approach mentionned above. In the `rest frame' $(X_a^0 = c \, T_a , X_a^i)$ attached to the pulsar $a$, the pulsar phenomenon can be modelled by the secularly changing rotation of a beam of radio waves:
\begin{equation}
\label{eq3.7}
\Phi_a = \int \Omega_a (T_a) \, d \, T_a \simeq \Omega_a \, T_a + \frac{1}{2} \, \dot\Omega_a \, T_a^2 + \frac{1}{6} \, \ddot\Omega_a \, T_a^3 + \cdots \, ,
\end{equation}
where $\Phi_a$ is the longitude around the spin axis. [Depending on the precise definition of the rest-frame attached to the pulsar, the spin axis can either be fixed, or be slowly evolving, see e.g. \cite{DSX}.] One must then relate the initial direction $(\Theta_a , \Phi_a)$, and proper time $T_a$, of emission of the pulsar beam to the coordinate direction and coordinate time of the null geodesic representing the electromagnetic beam in the `global' coordinates $x^{\mu}$ used to describe the dynamics of the binary system [NB: the explicit orbital motion (\ref{eq3.1})--(\ref{eq3.5}) refers to such global coordinates $x^0 = ct$, $x^i$]. This is done by using the link (\ref{eq2.6}) in which $z_a^i$ denotes the global coordinates of the `center of mass' of the pulsar, $T_a$ the local (proper) time of the pulsar frame, and where, for instance
\begin{equation}
\label{eq3.8}
e_i^0 = \frac{v_i}{c} \left( 1+\frac{1}{2} \ \frac{\bm{v}^2}{c^2} + 3 \, \frac{Gm_b}{c^2 \, r_{ab}} + \cdots \right) + \cdots
\end{equation}

Using the link (\ref{eq2.6}) (with expressions such as (\ref{eq3.8}) for the coefficients $e_i^{\mu} , \ldots$) one finds, among other results, that a radio beam emitted in the proper direction $N^i$ in the local frame appears to propagate, in the global frame, in the coordinate direction $n^i$ where
\begin{equation}
\label{eq3.9}
n^i = N^i + \frac{v^i}{c} - N^i \, \frac{N^j \, v^j}{c} + {\mathcal O} \left( \frac{v^2}{c^2} \right) \, .
\end{equation}

This is the well known `aberration effect', which will then contribute to the timing formula.

\smallskip

One must also write the link between the pulsar `proper time' $T_a$ and the coordinate time $t = x^0 / c = z_a^0 / c$ used in the orbital motion (\ref{eq3.1})--(\ref{eq3.5}). This reads
\begin{equation}
\label{eq3.10}
-c^2 \, d \, T_a^2 = \tilde g_{\mu\nu} (a_a^{\lambda}) \, dz_a^{\mu} \, dz_a^{\nu}
\end{equation}
where the `tilde' denotes the operation consisting (in the matching approach) in discarding in $g_{\mu\nu}$ the `self contributions' $\sim (Gm_a / R_a)^n$, while keeping the effect of the companion ($\sim Gm_b / r_{ab}$, etc$\ldots$). One checks that this is equivalent (in the dimensional-continuation approach) in taking $x^{\mu} = z_a^{\mu}$ for sufficiently {\it small} values of the real part of the dimension $d$. To lowest order this yields the link
\begin{equation}
\label{eq3.11}
T_a \simeq \int dt \left( 1 - \frac{2 \, Gm_b}{c^2 \, r_{ab}} - \frac{\bm{v}_a^2}{c^2} \right)^{\frac{1}{2}} \simeq \int dt  \left( 1 - \frac{Gm_b}{c^2 \, r_{ab}} - \frac{1}{2} \ \frac{\bm{v}_a^2}{c^2} \right)
\end{equation}
which combines the special relativistic and general relativistic time dilation effects. Hence, following \cite{DD86} we can refer to them as the `Einstein time delay'.

\smallskip

Then, one must compute the (global) time taken by a light beam emitted by the pulsar, at the proper time $T_a$ (linked to $t_{\rm emission}$ by (\ref{eq3.11})), in the initial global direction $n^i$ (see Eq.~(\ref{eq3.9})), to reach the barycenter of the solar system. This is done by writing that this light beam follows a null geodesic: in particular 
\begin{equation}
\label{eq3.12}
0 = ds^2 = g_{\mu\nu} (x^{\lambda}) \, dx^{\mu} \, dx^{\nu} \simeq - \left( 1-\frac{2U}{c^2} \right) c^2 \, dt^2 + \left( 1+\frac{2U}{c^2} \right) d\bm{x}^2
\end{equation}
where $U = Gm_a / \vert \bm{x} - \bm{z}_a \vert + Gm_b / \vert \bm{x} - \bm{z}_b \vert$ is the Newtonian potential within the binary system. This yields (with $t_e \equiv t_{\rm emission}$, $t_a \equiv t_{\rm arrival}$)
\begin{equation}
\label{eq3.13}
t_a - t_e = \int_{t_e}^{t_a} dt \simeq \frac{1}{c}  \int_{t_e}^{t_a} \vert d\bm{x} \vert + \frac{2}{c^3}  \int_{t_e}^{t_a}  \left( \frac{Gm_a}{\vert \bm{x} - \bm{z}_a \vert} + \frac{Gm_b}{\vert \bm{x} - \bm{z}_b \vert} \right) \vert d\bm{x} \vert \, .
\end{equation}

The first term on the last RHS of Eq.~(\ref{eq3.13}) is the usual `light crossing time' $\frac{1}{c} \, \vert \bm{z}_{\rm barycenter} (t_a) - \bm{z}_a (t_e) \vert$ between the pulsar and the solar barycenter. It contains the `Roemer time delay' due to the fact that $\bm{z}_a (t_e)$ moves on an orbit. The second term on the last RHS of Eq.~(\ref{eq3.13}) is the `Shapiro time delay' due to the propagation of the beam in a curved spacetime (only the $Gm_b$ piece linked to the companion is variable).

\smallskip

When inserting the `quasi-Keplerian' form (\ref{eq3.1})--(\ref{eq3.5}) of the relativistic motion in the `Roemer' term in (\ref{eq3.13}), together with all other relativistic effects, one finds that the final expression for the relativistic timing formula can be significantly simplified by doing two mathematical transformations. One can redefine the `time eccentricity' $e_t$ appearing in the `Kepler equation' (\ref{eq3.1}), and one can define a new `eccentric anomaly' angle: $u \to u^{\rm new}$ [we henceforth drop the superscript `new' on $u$]. After these changes, the binary-system part of the general relativistic timing formula \cite{DD86} takes the form (we suppress the index $a$ on the pulsar proper time $T_a$)
\begin{equation}
\label{eq3.14}
t_{\rm barycenter} - t_0 = D^{-1} [T + \Delta_R (T) + \Delta_E (T) + \Delta_S (T) + \Delta_A (T)]
\end{equation}
with
\begin{eqnarray}
\label{eq3.15}
\Delta_R &=  &x \sin \omega [\cos u - e (1 + \delta_r)] + x [1 - e^2 (1+\delta_{\theta})^2]^{1/2} \cos \omega \sin u \, , \\
\label{eq3.16}
\Delta_E &= &\gamma \sin u \, , \\
\label{eq3.17}
\Delta_S &=  &-2r \ln \{ 1 - e \cos u - s [\sin \omega (\cos u - e) + (1 - e^2)^{1/2} \cos \omega \sin u ]\} \!, \\
\label{eq3.18}
\Delta_A &= &A \{ \sin [\omega + A_e (u)] + e \sin \omega\} + B \{ \cos [\omega + A_e (u)] + e \cos \omega \} \, ,
\end{eqnarray}
where $x = x_0 + \dot x (T-T_0)$ represents the projected light-crossing time ($x = a_{\rm pulsar} \sin i / c$), $e = e_0 + \dot e (T-T_0)$ a certain (relativistically-defined) `timing eccentricity', $A_e (u)$ the function
\begin{equation}
\label{eq3.19}
A_e (u) \equiv 2 \, \arctan \left[ \left( \frac{1+e}{1-e} \right)^{1/2} \tan \frac{u}{2} \right] \, ,
\end{equation}
$\omega = \omega_0 + k \, A_e (u)$ the `argument of the periastron', and where the (relativisti\-cally-defined) `eccentric anomaly' $u$ is the function of the `pulsar proper time' $T$ obtained by solving the Kepler equation
\begin{equation}
\label{eq3.20}
u - e \sin u = 2\pi \left[ \frac{T - T_0}{P_b} - \frac{1}{2} \, \dot P_b \left( \frac{T - T_0}{P_b} \right)^2 \right] \,.
\end{equation}
It is understood here that the pulsar proper time $T$ corresponding to the $N^{\rm th}$ pulse is related to the integer $N$ by an equation of the form
\begin{equation}
\label{eq3.21}
N = c_0 + \nu_p \, T + \frac{1}{2} \, \dot\nu_p \, T^2 + \frac{1}{6} \, \ddot\nu_p \, T^3 \, .
\end{equation}
From these formulas, one sees that $\delta_{\theta}$ (and $\delta_r$) measure some relativistic distortion of the pulsar orbit, $\gamma$ the amplitude of the `Einstein time delay'\footnote{The post-Keplerian timing parameter $\gamma$, first introduced in \cite{BT76}, has the dimension of time, and should not be confused with the dimensionless post-Newtonian Eddington parameter $\gamma^{\rm
PPN}$ probed by solar-system experiments (see below).} $\Delta_E$, and $r$ and $s$ the {\it range} and {\it shape} of the `Shapiro time delay'\footnote{The dimensionless parameter $s$ is numerically equal to the sine of the inclination angle $i$ of the orbital plane, but its real definition within the PPK formalism is the
timing parameter which determines the `shape' of the logarithmic time delay $\Delta_S (T)$.} $\Delta_S$. Note also that the dimensionless PPK parameter $k$ measures the {\it non-uniform} advance of the periastron. It is related to the often quoted {\it secular} rate of periastron advance $\dot\omega \equiv \langle d\omega / dt \rangle$ by the relation $k = \dot\omega P_b / 2\pi$. It has been explicitly checked that binary-pulsar observational data do indeed require to model the relativistic periastron advance by means of the non-uniform (and non-trivial) function of $u$ multiplying $k$ on the R.H.S. of Eq.~(\ref{eq3.19}) \cite{Damour-Taylor92}\footnote{Alas this function is theory-independent, so that the non-uniform aspect of the periastron advance cannot be used to yield discriminating tests of
relativistic gravity theories.}. Finally, we see from Eq.~(\ref{eq3.20}) that $P_b$ represents the (periastron to periastron) orbital period at the fiducial epoch $T_0$, while the dimensionless parameter $\dot P_b$ represents the time derivative of $P_b$ (at $T_0$).

\smallskip

Schematically, the structure of the DD timing formula (\ref{eq3.14}) is
\begin{equation}
\label{eq3.22}
t_{\rm barycenter} - t_0 = F \, [T_N ; \{ p^K \} ; \{ p^{PK}\} ; \{ q^{PK}\}] \, ,
\end{equation}
where $t_{\rm barycenter}$ denotes the solar-system barycentric (infinite frequency) arrival time of a pulse, $T$ the pulsar emission proper time (corrected for aberration), $\{ p^K \} = \{ P_b , T_0 , e_0 , \omega_0 , x_0 \}$ is the set of {\it Keplerian} parameters, $\{ p^{PK} = k , \gamma , \dot P_b , r, s, \delta_{\theta} , \dot e , \dot x \}$ the set of {\it separately measurable post-Keplerian} parameters, and $\{ q^{PK} \} = \{ \delta_r , A, B, D \}$ the set of {\it not separately measurable post-Keplerian} parameters \cite{Damour-Taylor92}. [The parameter $D$ is a `Doppler factor' which enters as an overall multiplicative factor $D^{-1}$ on the right-hand side of Eq.~(\ref{eq3.14}).]

\smallskip

A further simplification of the DD timing formula was found possible. Indeed, the fact that the parameters $\{q^{PK}\} = \{ \delta_r , A,B,D \}$ are not separately measurable means that they can be absorbed in changes of the other parameters. The explicit formulas for doing that were given in \cite{DD86} and \cite{Damour-Taylor92}: they consist in redefining $e,x,P_b,\delta_{\theta}$ and $\delta_r$. At the end of the day, it suffices to consider a simplified timing formula where $\{ \delta_r , A , B , D\}$ have been set to some given fiducial values, e.g. $\{ 0,0,0,1 \}$, and where one only fits for the remaining parameters $\{ p^K \}$ and $\{p^{PK}\}$.

\smallskip

Finally, let us mention that it is possible to extend the general parametrized timing formula (\ref{eq3.22}) by writing a similar parametrized formula describing the effect of the pulsar orbital motion on the directional spectral luminosity $[d({\rm energy})$ $/ d({\rm time}) \, d({\rm frequency}) \, d(\mbox{solid angle})]$ received by an observer. As discussed in detail in \cite{Damour-Taylor92} this introduces a new set of `pulse-structure post-Keplerian parameters'.

\section{Phenomenological approach to testing relativistic gravity with binary pulsar data}\label{sec:4}

As said in the Introduction, binary pulsars contain strong gravity domains and should therefore allow one to test the strong-field aspects of relativistic gravity. The question we face is then the following: How can one use binary pulsar data to test strong-field (and radiative) gravity?

\smallskip

Two different types of answers can be given to this question: a {\it phenomenological} (or {\it theory-independent}) one, or various types of {\it theory-dependent} approaches. In this Section we shall consider the phenomenological approach.

\smallskip

The phenomenological approach to binary-pulsar tests of relativistic gravity is called the {\it parametrized post-Keplerian formalism} \cite{DamourCanada1988,Damour-Taylor92}. This approach is based on the fact that the mathematical form of the multi-parameter DD timing formula (\ref{eq3.22}) was found to be applicable not only in General Relativity, but also in a wide class of alternative theories of gravity. Indeed, any theory in which gravity is mediated not only by a metric field $g_{\mu\nu}$ but by a general combination of a metric field and of one or several scalar fields $\varphi^{(a)}$ will induce relativistic timing effects in binary pulsars which can still be parametrized by the formulas (\ref{eq3.14})--(\ref{eq3.21}). Such general `tensor-multi-scalar' theories of gravity contain arbitrary functions of the scalar fields. They have been studied in full generality in \cite{DEF92}. It was shown that, under certain conditions, such tensor-scalar gravity theories could lead, because of strong-field effects, to very different predictions from those of General Relativity in binary pulsar timing observations \cite{DEF93,DEF96,DEF98}. However, the point which is important for this Section, is that even when such strong-field effects develop one can still use the universal DD timing formula (\ref{eq3.22}) to fit the observed pulsar times of arrival.

\smallskip

The basic idea of the phenomenological, parametrized post-Keplerian (PPK) approach is then the following: By least-square fitting the observed sequence of pulsar arrival times $t_N$ to the parametrized formula (\ref{eq3.22}) (in which $T_N$ is defined by Eq.~(\ref{eq3.21}) which introduces the further parameters $\nu_p , \dot\nu_p , \ddot\nu_p$) one can phenomenologically extract from raw observational data the (best fit) values of all the parameters entering Eqs.~(\ref{eq3.21}) and (\ref{eq3.22}). In particular, one so determines both the set of Keplerian parameters $\{ p^K \} = \{ P_b , T_0 , e_0 , \omega_0 , x_0 \}$, and the set of post-Keplerian (PK) parameters $\{ p^{PK} \} = \{ k,\gamma, \dot P_b ,r,s,\delta_{\theta} , \dot e , \dot x \}$. In extracting these values, we did not have to assume any theory of gravity. However, each specific theory of gravity will make specific predictions relating the PK parameters to the Keplerian ones, and to the two (a priori unknown) masses $m_a$ and $m_b$ of the pulsar and its companion. [For certain PK parameters one must also consider other variables related to the spin vectors of $a$ and $b$.] In other words, the measurement (in addition of the Keplerian parameters) of each PK parameter defines, for each given theory, a {\it curve in the $(m_a , m_b)$ mass plane}. For any given theory, the measurement of two PK parameters determines two curves and thereby generically determines the values of the two masses $m_a$ and $m_b$ (as the point of intersection of these two curves). Therefore, as soon as one measures {\it three} PK parameters one obtains a {\it test} of the considered gravity theory. The test is passed only if the three curves meet at one point. More generally, the measurement of $n$ PK timing parameters yields $n-2$ independent tests of relativistic gravity. Any one of these tests, i.e. any simultaneous measurement of three PK parameters can either confirm or put in doubt any given theory of gravity.

\smallskip

As General Relativity is our current most successful theory of gravity, it is clearly the prime target for these tests. We have seen above that the timing data of each binary pulsar provides a maximum of 8 PK parameters: $k,\gamma , \dot P_b , r , s, \delta_{\theta} , \dot e$ and $\dot x$. Here, we were talking about a normal `single line' binary pulsar where, among the two compact objects $a$ and $b$ only one of the two, say $a$ is observed as a pulsar. In this case, one binary system can provide up to $8-2=6$ tests of GR. In practice, however, it has not yet been possible to measure the parameter $\delta_{\theta}$ (which measures a small relativistic deformation of the elliptical orbit), nor the secular parameters $\dot e$ and $\dot x$. The original Hulse-Taylor system PSR~1913$+$16 has allowed one to measure 3 PK parameters: $k \equiv \langle \dot\omega \rangle P_b / 2\pi$, $\gamma$ and $\dot P_b$. The two parameters $k$ and $\gamma$ involve (non radiative) strong-field effects, while, as explained above, the orbital period derivative $\dot P_b$ is a direct consequence of the term $A_5 \sim v^5/c^5$ in the binary-system equations of motion (\ref{eq2.5}). The term $A_5$ is itself directly linked to the retarded propagation, at the velocity of light, of the gravitational interaction between the two strongly self-gravitating bodies $a$ and $b$. Therefore, any test involving $\dot P_b$ will be a {\it mixed radiative strong-field} test.

\smallskip

Let us explain on this example what information one needs to implement a phenomenological test such as the $(k-\gamma - \dot P_b)_{1913+16}$ one. First, we need to know the predictions made by the considered target theory for the PK parameters $k,\gamma$ and $\dot P_b$ as functions of the two masses $m_a$ and $m_b$. These predictions have been worked out, for General Relativity, in Refs.~\cite{BT76,DamourPRL83,DD86}. Introducing the notation (where $n \equiv 2\pi/P_b$)
\begin{eqnarray}
\label{eq4.1}
M &\equiv &m_a + m_b \\
\label{eq4.2}
X_a &\equiv &m_a / M \, ; \quad X_b \equiv m_b/M \, ; \quad X_a + X_b \equiv 1 \\
\label{eq4.3}
\beta_O (M) &\equiv &\left( \frac{GMn}{c^3} \right)^{1/3} \, , 
\end{eqnarray}
they read
\begin{eqnarray}
\label{eq4.4}
k^{\rm GR} (m_a , m_b) &= &\frac{3}{1-e^2} \ \beta_O^2 \, , \\
\label{eq4.5}
\gamma^{\rm GR} (m_a , m_b) &= &\frac{e}{n} \ X_b (1+X_b) \, \beta_O^2 \, , \\
\label{eq4.6}
\dot P_b^{\rm GR} (m_a , m_b) &= &- \frac{192 \pi}{5} \ \frac{1 + \frac{73}{24} \, e^2 + \frac{37}{96} \, e^4}{(1-e^2)^{7/2}} \ X_a \, X_b \, \beta_O^5 \, .
\end{eqnarray}

However, if we use the three predictions (\ref{eq4.4})--(\ref{eq4.6}), together with the best current observed values of the PK parameters $k^{\rm obs} , \gamma^{\rm obs} , \dot P_b^{\rm obd}$ \cite{WeisbergTaylor04} we shall find that the three curves $k^{\rm GR} (m_a , m_b) = k^{\rm obs}$, $\gamma^{\rm GR} (m_a , m_b) = \gamma^{\rm obs}$, $\dot P_b^{\rm GR} (m_a , m_b) = \dot P_b^{\rm obs}$ in the $(m_a , m_b)$ mass plane {\it fail to meet} at about the $13 \, \sigma$ level! Should this put in doubt General Relativity? No, because Ref.~\cite{DamourTaylor91} has shown that the time variation (notably due to galactic acceleration effects) of the Doppler factor $D$ entering Eq.~(\ref{eq3.14}) entailed an extra contribution to the `observed' period derivative $\dot P_b^{\rm obs}$. We need to subtract this non-GR contribution before drawing the corresponding curve: $\dot P_b^{\rm GR} (m_a , m_b) = \dot P_b^{\rm obs} - \dot P_b^{\rm galactic}$. Then one finds that the three curves {\it do meet} within one $\sigma$. This yields a deep confirmation of General Relativity, and a direct observational proof of the reality of gravitational radiation.

\smallskip

We said several times that this test is also a probe of the strong-field aspects of GR. How can one see this? A look at the GR predictions (\ref{eq4.4})--(\ref{eq4.6}) does not exhibit explicit strong-field effects. Indeed, the derivation of Eqs.~(\ref{eq4.4})--(\ref{eq4.6}) used in a crucial way the `effacement of internal structure' that occurs in the general relativistic dynamics of compact objects. This non trivial property is rather specific of GR and means that, in this theory, all the strong-field effects can be {\it absorbed} in the definition of the masses $m_a$ and $m_b$. One can, however, verify that strong-field effects do enter the observable PK parameters $k,\gamma , \dot P_b$ etc$\ldots$ by considering how the theoretical predictions (\ref{eq4.4})--(\ref{eq4.6}) get modified in alternative theories of gravity. The presence of such strong-field effects in PK parameters was first pointed out in Ref.~\cite{Eardley75} (see also \cite{WillZaglauer89}) for the Jordan-Fierz-Brans-Dicke theory of gravity, and in Ref.~\cite{WillEardley77} for Rosen's bi-metric theory of gravity. A detailed study of such strong-field deviations was then performed in \cite{DEF92,DEF93,DEF96} for general tensor-(multi-)scalar theories of gravity. In the following Section we shall exhibit how such strong-field effects enter the various post-Keplerian parameters.

\smallskip

Continuing our historical review of phenomenological pulsar tests, let us come to the binary system which was the first one to provide several `pure strong-field tests' of relativistic gravity, without mixing of radiative effects: PSR~1534$+$12. In this system, it was possible to measure the four (non radiative) PK parameters $k,\gamma,r$ and $s$. [We see from Eq.~(\ref{eq3.17}) that $r$ and $s$ measure, respectively, the {\it range} and the {\it shape} of the `Shapiro time delay' $\Delta_S$.] The measurement of the 4 PK parameters $k,\gamma,r,s$ define 4 curves in the $(m_a , m_b)$ mass plane, and thereby yield 2 strong-field tests of GR. It was found in \cite{TWDW92} that GR passes these two tests. For instance, the ratio between the measured value $s^{\rm obs}$ of the phenomenological parameter\footnote{As already mentioned the dimensionless parameter $s$ is numerically equal (in all theories) to the sine of the inclination angle $i$ of the orbital plane, but it is better thought, in the PPK formalism, as a phenomenological timing parameter determining the `shape' of the logarithmic time delay $\Delta_S (T)$.} $s$ and the value $s^{\rm GR} [k^{\rm obs} , \gamma^{\rm obs}]$ predicted by GR on the basis of the measurements of the two PK parameters $k$ and $\gamma$ (which determine, via  Eqs.~(\ref{eq4.4}) , (\ref{eq4.5}), the GR-predicted value of $m_a$ and $m_b$) was found to be $s^{\rm obs} / s^{\rm GR} [k^{\rm obs} , \gamma^{\rm obs}] = 1.004 \pm 0.007$ \cite{TWDW92}. The most recent data \cite{Stairs02} yield $s^{\rm obs} / s^{\rm GR} [k^{\rm obs} , \gamma^{\rm obs}] = 1.000 \pm 0.007$. We see that we have here a confirmation of the strong-field regime of GR at the 1\% level.

\smallskip

Another way to get phenomenological tests of the strong field aspects of gravity concerns the possibility of a violation of the strong equivalence principle. This is parametrized by phenomenologically assuming that the ratio between the gravitational and the inertial mass of the pulsar differs from unity (which is its value in GR): $(m_{\rm grav} / m_{\rm inert})_a = 1 + \Delta_a$. Similarly to what happens in the Earth-Moon-Sun system \cite{Nordtvedt68}, the three-body system made of a binary pulsar and of the Galaxy exhibits a `polarization' of the orbit which is proportional to $\Delta \equiv \Delta_a - \Delta_b$, and which can be constrained by considering certain quasi-circular neutron-star-white-dwarf binary systems \cite{DS91}. See \cite{Stairsetal} for recently published improved limits\footnote{Note, however, that these limits, as well as those previously obtained in \cite{Wex97}, assume that the (a priori pulsar-mass dependent) parameter $\Delta \simeq \Delta_a$ is the same for all the analyzed pulsars.} on the phenomenological equivalence-principle violation parameter $\Delta$.

\smallskip

The Parkes multibeam survey has recently discovered several new interesting `relativistic' binary pulsars, thereby giving a huge increase in the number of phenomenological tests of relativistic gravity. Among those new binary pulsar systems, two stand out as superb testing grounds for relativistic gravity: (i) PSR J1141$-$6545 \cite{Kaspietal,Bailes03}, and (ii) the remarkable double binary pulsar PSR J0737$-$3039A and B \cite{Burgay03,Lyne04,Kramer04,Krameretal06} (see also the lectures by M.~Kramer and A.~Possenti).

\smallskip

The PSR J1141$-$6545 timing data have led to the measurement of 3 PK parameters: $k$, $\gamma$, and $\dot P_b$ \cite{Bailes03}. As in PSR 1913$+$16 this yields one mixed radiative-strong-field test\footnote{In addition, scintillation data have led to an estimate of the sine of the orbital inclination, $\sin i$ \cite{Ord02}. As said above, $\sin i$ numerically coincides with the PK parameter $s$ measuring the `shape' of the Shapiro time delay. Therefore, one could use the scintillation measurements as an indirect determination of $s$, thereby obtaining two independent tests from PSR J1141$-$6545 data. A caveat, however, is that the extraction of $\sin i$ from scintillation measurements rests on several simplifying assumptions whose validity is unclear. In fact, in the case of PSR J0737$-$3039 the direct timing measurement of $s$ disagrees with its estimate via scintillation data \cite{Krameretal06}. It is therefore safer not to use scintillation estimates of $\sin i$ on the same footing as direct timing measurements of the PK parameter $s$. On the other hand, a safe way of obtaining an $s$-related gravity test consists in using the necessary mathematical fact that $s = \sin i \leq 1$. In GR the definition $x_a = a_a \sin i / c$ leads to $\sin i = n \, x_a / (\beta_0 \, X_b)$. Therefore we can write the inequality $n \, x_a / (\beta_0 (M) \, X_b) \leq 1$ as a phenomenological test of GR.}.

\smallskip

The timing data of the millisecond binary pulsar PSR J0737$-$3039A have led to the direct measurement of 5 PK parameters: $k$, $\gamma$, $r$, $s$ and $\dot P_b$ \cite{Lyne04,Kramer04,Krameretal06}. In addition, the `double line' nature of this binary system (i.e. the fact that one observes both components, $A$ {\it and} $B$, as radio pulsars) allows one to perform new phenomenological tests by using {\it Keplerian} parameters. Indeed, the simultaneous measurement of the Keplerian parameters $x_a$ and $x_b$ representing the projected light crossing times of both pulsars ($A$ and $B$) gives access to the combined Keplerian parameter
\begin{equation}
\label{eq4.7}
R^{\rm obs} \equiv \frac{x_b^{\rm obs}}{x_a^{\rm obs}} \, .
\end{equation}

On the other hand, the general derivation of \cite{DD86} (applicable to any Lorentz-invariant theory of gravity, and notably to any tensor-scalar theory) shows that the theoretical prediction for the the ratio $R$, considered as a function of the masses $m_a$ and $m_b$, is
\begin{equation}
\label{eq4.8}
R^{\rm theory} = \frac{m_a}{m_b} + {\mathcal O} \left( \frac{v^4}{c^4} \right) \, .
\end{equation}
The absence of any {\it explicit} strong-field-gravity effects in the theoretical prediction (\ref{eq4.8}) (to be contrasted, for instance, with the predictions for PK parameters in tensor-scalar gravity discussed in the next Section) is mainly due to the convention used in \cite{DD86} and \cite{Damour-Taylor92} for {\it defining} the masses $m_a$ and $m_b$. These are always defined so that the Lagrangian for two non interacting compact objects reads $L_0 = \underset{a}{\sum} - m_a \, c^2 (1-\bm{v}_a^2 / c^2)^{1/2}$. In other words, $m_a \, c^2$ represents the total energy of body $a$. This means that one has {\it implicitly} lumped in the definition of $m_a$ many strong-self-gravity effects. [For instance, in tensor-scalar gravity $m_a$ includes not only the usual Einsteinian gravitational binding energy due to the self-gravitational field $g_{\mu\nu} (x)$, but also the extra binding energy linked to the scalar field $\varphi (x)$.] Anyway, what is important is that, when performing a phenomenological test from the measurement of a triplet of parameters, e.g. $\{ k , \gamma , R \}$, {\it at least one} parameter among them be a priori sensitive to strong-field effects. This is enough for guaranteeing that the crossing of the three curves $k^{\rm theory} (m_a , m_b) = k^{\rm obs}$, $\gamma^{\rm theory} (m_a , m_b) = \gamma^{\rm obs}$, $R^{\rm theory} (m_a , m_b) = R^{\rm obs}$ is really a probe of strong-field gravity.

%
%
\begin{figure}[h]
\centering
\includegraphics[height=8cm]{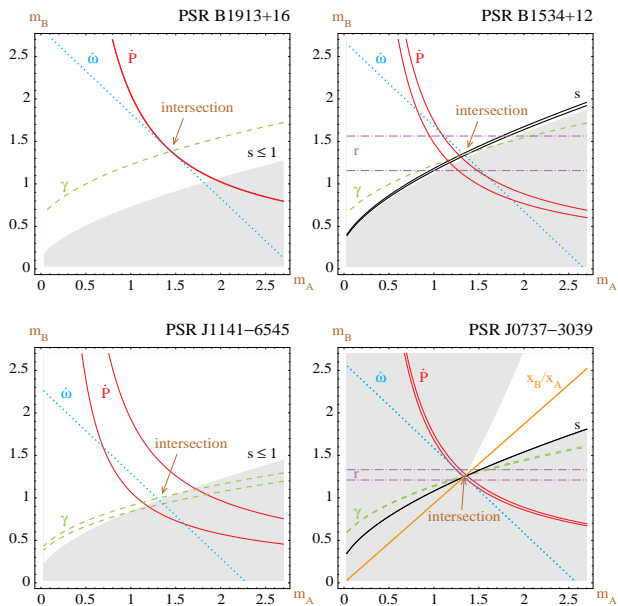}
%
%
\caption{Phenomenological tests of General Relativity obtained from Keplerian and post-Keplerian timing parameters of four relativistic pulsars. Figure taken from \cite{DEF06}.}
\label{fig:1}       
\end{figure}

In conclusion, the two recently discovered binary pulsars PSR J1141$-$6545 and PSR J0737$-$3039 have more than {\it doubled} the number of phenomenological tests of (radiative and) strong-field gravity. Before their discovery, the `canonical' relativistic binary pulsars PSR 1913$+$16 and PSR 1534$+$12 had given us {\it four} such tests: one $(k - \gamma - \dot P_b)$ test from PSR 1913$+$16 and three ($k-\gamma -r-s-\dot P_b$\footnote{The timing measurement of $\dot P_b^{\rm obs}$ in PSR 1534$+$12 is even more strongly affected by kinematic corrections ($\dot D$ terms) than in the PSR 1913$+$16 case. In absence of a precise, independent measurement of the distance to PSR 1534$+$12, the $k-\gamma - \dot P_b$ test yields, at best, a $\sim$ 15\% test of GR.}) tests from PSR 1534$+$12. The two new binary systems have given us {\it five}\footnote{Or even six, if we use the scintillation determination of $s$ in PSR J1141$-$6545.} more phenomenological tests: one $(k-\gamma - \dot P_b)$ (or two, $k-\gamma - \dot P_b - s$) tests from PSR J1141$-$6545 and four ($k-\gamma -r-s-\dot P_b - R$) tests from PSR J0737$-$3039\footnote{The companion pulsar 0737$-$3039B being non recycled, and being visible only during a small part of its orbit, cannot be timed with sufficient accuracy to allow one to measure any of its post-Keplerian parameters.}. As illustrated in Figure~\ref{fig:1}, these {\it nine} phenomenological tests of strong-field (and radiative) gravity are all in beautiful agreement with General Relativity.

\smallskip

In addition, let us recall that several quasi-circular wide binaries, made of a neutron star and a white dwarf, have led to high-precision phenomenological confirmations \cite{Stairsetal} (in strong-field conditions) of one of the deep predictions of General Relativity: the `strong' equivalence principle, i.e. the fact that various bodies fall with the same acceleration in an external gravitational field, independently of the strength of their self-gravity.

\smallskip

Finally, let us mention that Ref.~\cite{Damour-Taylor92} has extended the philosophy of the phenomenological (parametrized post-Keplerian) analysis of timing data, to a similar phenomenological analysis of {\it pulse-structure data}. Ref.~\cite{Damour-Taylor92} showed that, in principle, one could extract up to 11 `post-Keplerian pulse-structure parameters'. Together with the $8$ post-Keplerian timing parameters of a (single-line) binary pulsar, this makes a total of $19$ phenomenological PK parameters. As these parameters depend not only on the two masses $m_a , m_b$ but also on the two angles $\lambda , \eta$ determining the direction of the spin axis of the pulsar, the maximum number of tests one might hope to extract from one (single-line) binary pulsar is $19-4 = 15$. However, the present accuracy with which one can model and measure the pulse structure of the known pulsars has not yet allowed one to measure any of these new pulse-structure parameters in a theory-independent and model-independent way. 

\smallskip

Nonetheless, it has been possible to confirm the reality (and order of magnitude) of the spin-orbit coupling in GR which was pointed out \cite{DamourRuffini74,BarkerOConnel75} to be observable via a secular change of the intensity profile of a pulsar signal. Confirmations of general relativistic spin-orbit effects in the evolution of pulsar profiles were obained in several pulsars: PSR 1913$+$16 \cite{Kramer98,WeisbergTaylor02}, PSR B1534$+$12 \cite{StairsThorsettArzoumanian04} and PSR J1141$-$6545 \cite{HotanBailesOrd05}. In this respect, let us mention that the spin-orbit interaction affects also several PK parameters, either by inducing a secular evolution in some of them (see \cite{Damour-Taylor92}) or by contributing to their value. For instance, the spin-orbit interaction contributes to the observed value of the periastron advance parameter $k$ an amount which is significant for the pulsars (such as 1913$+$16 and 0737$-$3039) where $k$ is measured with high-accuracy. It was then pointed out \cite{DS88} that this gives, in principle, and indirect way of measuring the moment of inertia of neutron stars (a useful quantity for probing the equation of state of nuclear matter \cite{LattinerSchutz05,Morrisonetal04}). However, this can be done only if one measures, besides $k$, {\it two other} PK parameters with $10^{-5}$ accuracy. A rather tall order which will be a challenge to meet.

\smallskip

The phenomenological approach to pulsar tests has the advantage that it can confirm or invalidate a specific theory of gravity without making assumptions about other theories. Moreover, as General Relativity has no free parameters, any test of its predictions is a potentially lethal test. From this point of view, it is remarkable that GR has passed with flying colours all the pulsar tests if has been submitted to. [See, notably, Fig.~\ref{fig:1}.] As argued above, these tests have probed strong-field aspects of gravity which had not been probed by solar-system (or cosmological) tests. On the other hand, a disadvantage of the phenomenological tests is that they do not tell us in any precise way which strong-field structures, have been actually tested. For instance, let us imagine that one day one specific PPK test fails to be satisfied by GR, while the others are OK. This leaves us in a quandary: If we trust the problematic test, we must conclude that GR is wrong. However, the other tests say that GR is OK. This example shows that we would like to have some idea of what physical effects, linked to strong-field gravity, enter in each test, or even better in each PK parameter. The `effacement of internal structure' which takes place in GR does not allow one to discuss this issue. This gives us a motivation for going beyond the phenomenological PPK approach by considering {\it theory-dependent} formalisms in which one embeds GR within a {\it space of alternative gravity theories}.

\section{Theory-space approach to testing relativistic gravity with binary pulsar data}\label{sec:5}

A complementary approach to testing gravity with binary pulsar data consists in embedding General Relativity within a {\it multi-parameter space of alternative theories of gravity}. In other words, we want to {\it contrast} the predictions of GR with the predictions of continuous families of alternative theories. In so doing we hope to learn more about which structures of GR are actually being probed in binary pulsar tests. This is a bit similar to the well-known psycho-physiological fact that the best way to appreciate a nuance of colour is to surround a given patch of colour by other patches with slightly different colours. This makes it much easier to detect subtle differences in colour. In the same way, we hope to learn about the probing power of pulsar tests by seeing how the phenomenological tests summarized in Fig.~\ref{fig:1} fail (or continue) to be satisfied when one continuously deform, away from GR, the gravity theory which is being tested.

\smallskip

Let us first recall the various ways in which this {\it theory-space approach} has been used in the context of the solar-system tests of relativistic gravity. 

\subsection{Theory-space approaches to solar-system tests of relativistic gravity}\label{ssec:5.1}

In the quasi-stationary weak-field context of the solar-system, this theory-space approach has been implemented in two different ways. First, the parametrized post-Newtonian (PPN) formalism \cite{Eddington23,Schiff60,Baierlein67,Nordtvedt68,Will71,WillNordtvedt72,Will93,Will01} describes many `directions' in which generic alternative theories of gravity might differ in their {\it weak-field} predictions from GR. In its most general versions the PPN formalism contains $10$ `post-Einstein' PPN parameters, $\bar\gamma \equiv \gamma^{\rm PPN} - 1$\footnote{The PPN parameter $\gamma^{\rm PPN}$ is usually denoted simply as $\gamma$. To distinguish it from the Einstein-time-delay PPK timing parameter $\gamma$ used above we add the superscript PPN. In addition, as the value of $\gamma^{\rm PPN}$ in GR is $1$, we prefer to work with the parameter $\bar\gamma \equiv \gamma^{\rm PPN} - 1$ which vanishes in GR, and therefore measures a `deviation' from GR in a certain `direction' in theory-space. Similarly with $\bar\beta \equiv \beta^{\rm PPN} - 1$.}, $\bar\beta \equiv \beta^{\rm PPN} - 1 , \xi , \alpha_1 , \alpha_2 , \alpha_3 , \zeta_1 , \zeta_2 , \zeta_3 , \zeta_4$. Each one of these dimensionless quantities parametrizes a certain class of slow-motion, weak-field gravitational effects which deviate from corresponding GR predictions. For instance, $\bar\gamma$ parametrizes modifications both of the effect of a massive body (say, the Sun) on the light passing near it, and of the terms in the two-body gravitational Lagrangian which are proportional to $(G \, m_a \, m_b / r_{ab}) \cdot (\bm{v}_a - \bm{v}_b)^2 / c^2$.

\smallskip

A second way of implementing the theory-space philosophy consists in considering some explicit, parameter-dependent family of alternative relativistic theories of gravity. For instance, the simplest tensor-scalar theory of gravity put forward by Jordan \cite{Jordan49}, Fierz \cite{Fierz56} and Brans and Dicke \cite{BransDicke61} has a unique free parameter, say $\alpha_0^2 = (2 \, \omega_{\rm BD} + 3)^{-1}$. When $\alpha_0^2 \to 0$, this theory reduces to GR, so that $\alpha_0^2$ (or $1/\omega_{\rm BD}$) measures all the deviations from GR. When considering the weak-field limit of the Jordan-Fierz-Brans-Dicke (JFBD) theory, one finds that it can be described within the PPN formalism by choosing $\bar\gamma = - 2 \, \alpha_0^2 (1+\alpha_0^2)^{-1}$, $\bar\beta = 0$ and $\xi = \alpha_i = \zeta_j = 0$.

\smallskip

Having briefly recalled the two types of theory-space approaches used to discuss solar-system tests, let us now consider the case of binary-pulsar tests.

\subsection{Theory-space approaches to binary-pulsar tests of relativistic gravity}\label{ssec:5.2}

There exist generalizations of these two different theory-space approaches to the context of strong-field gravity and binary pulsar tests. First, the PPN formalism has been (partially) extended beyond the `first post-Newtonian' (1PN) order deviations from GR ($\sim v^2 / c^2 + Gm / c^2 \, r$) to describe 2PN order deviations from GR $\left( \sim \left( \frac{v^2}{c^2} + \frac{Gm}{c^2 \, r} \right)^2\right)$ \cite{DEF2PN}. Remarkably, there appear only two new parameters at the 2PN level\footnote{When restricting oneself to the general class of tensor-multi-scalar theories. At the 1PN level, this restriction would imply that only the `directions' $\bar\gamma$ and $\bar\beta$ are allowed.}: $\epsilon$ and $\zeta$. Also, by expanding in powers of the self-gravity parameters of body $a$ and $b$ the predictions for the PPK timing parameters in generic tensor-multi-scalar theories, one has shown that these predictions depended on several `layers' of new dimensionless parameters \cite{DEF92}. Early among these parameters one finds, the 1PN parameters $\bar\beta , \bar\gamma$ and then the basic 2PN parameters $\epsilon$ and $\zeta$, but one also finds further parameters $\beta_3$, $(\beta\beta')$, $\beta'' , \ldots$ which would not enter usual 2PN effects. The two approaches that we have just mentionned can be viewed as generalizations of the PPN formalism.

\smallskip

There exist also useful generalizations to the strong-field context of the idea of considering some explicit parameter-dependent family of alternative theories of relativistic gravity. Early studies \cite{Eardley75,WillEardley77,WillZaglauer89} focussed either on the one-parameter JFBD tensor-scalar theory, or on some theories which are not continuously connected to GR, such as Rosen's bimetric theory of gravity. Though the JFBD theory exhibits a marked difference from GR in that it predicts the existence of dipole radiation, it has the disadvantage that the weak field, solar-system constraints on its unique parameter $\alpha_0^2$ are so strong that they drastically constrain (and essentially forbid) the presence of any non-radiative, strong-field deviations from GR. In view of this, it is useful to consider other `mini-spaces' of alternative theories. 

\smallskip

A two-parameter mini-space of theories, that we shall denote\footnote{We add here an index 2 to $T$ as a reminder that this is a class of tensor-{\it bi}-scalar theories, i.e. that they contain {\it two} independent scalar fields $\varphi_1 , \varphi_2$ besides a dynamical metric $g_{\mu\nu}$.} here as $T_2 (\beta' , \beta'')$, was introduced in \cite{DEF92}. This two-parameter family of tensor-bi-scalar theories was constructed so as to have exactly the same first post-Newtonian limit as GR (i.e. $\bar\gamma = \bar\beta = \cdots = 0$), but to differ from GR in its predictions for the various observables that can be extracted from binary pulsar data. Let us give one example of this behaviour of the $T_2 (\beta' , \beta'')$ class of theories. For a general theory of gravity we expect to have violations of the strong equivalence principle in the sense that the ratio between the gravitational mass of a self-gravitating body to its inertial mass will admit an expansion of the type
\begin{equation}
\label{eq5.1}
\frac{m_a^{\rm grav}}{m_a^{\rm inert}} \equiv 1 + \Delta_a = 1 - \frac{1}{2} \, \eta_1 \, c_a + \eta_2 \, c_a^2 + \ldots
\end{equation}
where $c_a \equiv -2 \, \frac{\partial \ln m_a}{\partial \ln G}$ measures the `gravitational compactness' (or fractional gravitational binding energy, $c_a \simeq -2 \, E_a^{\rm grav}/ m_a \, c^2$) of body $a$. The numerical coefficient $\eta_1$ of the contribution linear in $c_a$ is a combination of the first post-Newtonian order PPN parameters, namely $\eta_1 = 4 \, \bar \beta - \bar \gamma$ \cite{Nordtvedt68}. The numerical coefficient $\eta_2$ of the term quadratic in $c_a$ is a combination of the 1PN {\it and} 2PN parameters. When working in the context of the $T_2 (\beta' , \beta'')$ theories, the 1PN parameters vanish exactly $(\bar\beta = 0 = \bar\gamma)$ and the coefficient of the quadratic term becomes simply proportional to the theory parameter $\beta' : \eta_2 = \frac{1}{2} \, B \beta'$, where $B \approx 1.026$. This example shows explicitly how binary pulsar data (here the data constraining the equivalence principle violation parameter $\Delta = \Delta_a - \Delta_b$, see above) can go beyond solar-system experiments in probing certain strong-self-gravity effects. Indeed, solar-system experiments are totally insensitive to 2PN parameters because of the smallness of $c_a \sim Gm_a / c^2 \, R_a$ and of the structure of 2PN effects \cite{DEF2PN}. By contrast, the `compactness' of neutron stars is of order $c_a \sim 0.21 \, m_a / M_{\odot} \sim 0.3$ \cite{DEF92} so that the pulsar limit $\vert \Delta \vert < 5.5 \times 10^{-3}$ \cite{Stairsetal} yields, within the $T_2 (\beta' , \beta'')$ framework, a significant limit on the dimensionless (2PN order) parameter $\beta' : \vert \beta' \vert < 0.12$.

\smallskip

Ref.~\cite{DEF96} introduced a new two-parameter mini-space of gravity theories, denoted here as $T_1 (\alpha_0 , \beta_0)$, which, from the point of view of theoretical physics, has several advantages over the $T_2 (\beta',\beta'')$ mini-space mentionned above. First, it is technically simpler in that it contains only {\it one} scalar field $\varphi$ besides the metric $g_{\mu\nu}$ (hence the index $1$ on $T_1 (\alpha_0 , \beta_0)$). Second, it contains only positive-energy excitations (while one combination of the two scalar fields of $T_2 (\beta' , \beta'')$ carried negative-energy waves). Third, it is the minimal way to parametrize the huge class of tensor-mono-scalar theories with a `coupling function' $a(\varphi)$ satisfying some very general requirements (see below).

\smallskip

Let us now motivate the use of tensor-scalar theories of gravity as alternatives to general relativity.

\subsection{Tensor-scalar theories of gravity}\label{ssec:5.3}

Let us start by recalling (essentially from \cite{DEF96}) why tensor-(mono)-scalar theories define a natural class of alternatives to GR.
First, and foremost, the existence of scalar partners to the graviton is a simple theoretical possibility which has surfaced many times in the development of unified theories, from Kaluza-Klein to superstring theory. Second, they are general enough to describe many interesting deviations from GR (both in weak-field and in strong field conditions), but simple enough to allow one to work out their predictions in full detail. 

\smallskip

Let us therefore consider a general tensor-scalar action involving a metric $\tilde g_{\mu\nu}$
(with signature `mostly plus'), a scalar field $\Phi$, and some matter variables $\psi_m$ (including gauge bosons):
\begin{equation}
\label{eq2.1:5.2}
S = \frac{c^4}{16 \pi \, G_*} \int \frac{d^4 x}{c} \ \tilde g^{1/2}
\left[ F(\Phi) \tilde R - Z (\Phi) \tilde g^{\mu\nu} \partial_{\mu}
\Phi \, \partial_{\nu} \Phi - U (\Phi) \right] + S_m [\psi_m ; \tilde
g_{\mu\nu}] \,.
\end{equation}
For simplicity, we assume here that the weak equivalence principle is satisfied, i.e., that the matter variables $\psi_m$ are all coupled to the same `physical metric'\footnote{Actually, most unified models
suggest that there are violations of the weak equivalence principle. However, the study of general string-inspired tensor-scalar models \cite{Damour-Polyakov94} has found that the composition-dependent
effects would be negligible in the gravitational physics of neutron stars that we consider here. The experimental limits on tests of the equivalence principle would, however, bring a strong additional
constraint of order $10^{-5} \, \alpha_0^2 \sim \Delta a/a \lesssim 10^{-12}$. As this constraint is strongly model-dependent, we will not use it in our exclusion plots below. One should, however, keep in mind
that a limit on the scalar coupling strength $\alpha_0^2$ of order $\alpha_0^2 \lesssim 10^{-7}$
\cite{Damour-Polyakov94,Damour-Vokrouhlicky96} is likely to exist in many, physically-motivated, tensor-scalar models.} $\tilde g_{\mu\nu}$. The general model (\ref{eq2.1:5.2}) involves three arbitrary
functions: a function $F(\Phi)$ coupling the scalar $\Phi$ to the Ricci scalar of $\tilde g_{\mu\nu}$, $\tilde R \equiv R (\tilde g_{\mu\nu})$, a function $Z(\Phi)$ renormalizing the kinetic term of $\Phi$, and a potential function $U(\Phi)$. As we have the freedom of arbitrary redefinitions of the scalar field, $\Phi \to \Phi' = f(\Phi)$, only two functions among $F$, $Z$ and $U$ are independent. It is often convenient to
rewrite (\ref{eq2.1:5.2}) in a {\it canonical} form, obtained by redefining both $\Phi$ and $\tilde g_{\mu\nu}$ according to
\begin{equation}
\label{eq2.2:5.3}
g_{\mu\nu}^* = F(\Phi) \, \tilde g_{\mu\nu} \, ,
\end{equation}
\begin{equation}
\label{eq2.3:5.4}
\varphi = \pm \int d\Phi \left[ \frac{3}{4} \, \frac{F'^2 (\Phi)}{F^2
(\Phi)} + \frac{1}{2} \, \frac{Z(\Phi)}{F(\Phi)} \right]^{1/2} \,.
\end{equation}
This yields
\begin{equation}
\label{eq2.4:5.5}
S = \frac{c^4}{16 \pi \, G_*} \int \frac{d^4 x}{c} \ g_*^{1/2} \left[
R_* - 2 g_*^{\mu\nu} \partial_{\mu} \varphi \, \partial_{\nu} \varphi
- V(\varphi) \right] + S_m \left[ \psi_m ; A^2 (\varphi) \,
g_{\mu\nu}^* \right] \, ,
\end{equation}
where $R_* \equiv R(g_{\mu\nu}^*)$, where the potential
\begin{equation}
\label{eq2.5:5.6}
V(\varphi) = F^{-2} (\Phi) \, U(\Phi) \, ,
\end{equation}
and where the {\it conformal coupling function} $A(\varphi)$ is given by
\begin{equation}
\label{eq2.6:5.7}
A(\varphi) = F^{-1/2} (\Phi) \, ,
\end{equation}
with $\Phi (\varphi)$ obtained by inverting the integral (\ref{eq2.3:5.4}).

\smallskip

The two arbitrary functions entering the canonical form (\ref{eq2.4:5.5}) are: (i) the conformal coupling function $A(\varphi)$, and (ii) the potential function $V(\varphi)$. Note that the `physical metric'
$\tilde g_{\mu\nu}$ (the one measured by laboratory clocks and rods) is conformally related to the `Einstein metric' $g_{\mu\nu}^*$, being given by $\tilde g_{\mu\nu} = A^2 (\varphi) \, g_{\mu\nu}^*$. The
canonical representation is technically useful because it decouples the two irreducible propagating excitations: the spin-0 excitations are described by $\varphi$, while the pure spin-2 excitations are
described by the {\it Einstein metric} $g_{\mu\nu}^*$ (with kinetic term the usual Einstein-Hilbert action $\propto R (g_{\mu\nu}^*)$).

\smallskip

In many technical developments it is useful to work with the {\it logarithmic coupling function} $a(\varphi)$ such that:
\begin{equation}
\label{aphi}
a(\varphi) \equiv \ln A(\varphi) \, ; \ A(\varphi) \equiv
e^{a(\varphi)} \,.
\end{equation}
In the case of the general model (\ref{eq2.1:5.2}) this logarithmic\footnote{As we shall mostly work with $a(\varphi)$ below, we shall henceforth drop the adjective `logarithmic'.} coupling function is given by
$$
a(\varphi) = -\frac{1}{2} \, \ln F(\Phi) \, ,
$$
where $\Phi (\varphi)$ must be obtained from (\ref{eq2.3:5.4}).

\smallskip

In the following, we shall assume that the potential $V(\varphi)$ is a slowly varying function of $\varphi$ which, in the domain of variation we shall explore, is roughly equivalent to a very small mass term $V(\varphi) \sim 2 \, m_{\varphi}^2 (\varphi - \varphi_0)^2$ with $m_{\varphi}^2$ of cosmological order of magnitude $m_{\varphi}^2 = {\mathcal O} (H_0^2)$, or, at least, with a range $\lambda_{\varphi} = m_{\varphi}^{-1}$ much larger than the typical length scales that we shall consider (such as the size of the binary orbit, or the size of the Galaxy when considering violations of the strong equivalence principle). Under this assumption\footnote{Note, however, that, as was recently explored in
\cite{Khoury-Weltman04,KW2,Brax04}, a sufficiently fast varying potential $V(\varphi)$ can change the tensor-scalar phenomenology by endowing $\varphi$ with a mass term $m_{\varphi}^2 = \frac{1}{4} \, \partial^2
V / \partial \varphi^2$ which strongly depends on the local value of $\varphi$ and, thereby can get large in sufficiently dense environments.} the potential function $V(\varphi)$ will only serve the role of fixing the value of $\varphi$ far from the system (to $\varphi (r=\infty) = \varphi_0$), and its effect on the propagation of $\varphi$ within the system will be negligible. In the end, the tensor-scalar phenomenology that we shall explore only depends on {\it one function}: the coupling function $a(\varphi)$.

\smallskip

Let us consider some examples to see what kind of coupling functions might naturally arise. First, the simplest case is the Jordan-Fierz-Brans-Dicke action, which is of the general type
(\ref{eq2.1:5.2}) with
\begin{eqnarray}
\label{eq2.7:5.8}
F(\Phi) &= & \Phi \\
\label{2.7bis:5.8bis}
Z(\Phi) &= & \omega_\text{BD} \, \Phi^{-1} \, ,
\end{eqnarray}
where $\omega_\text{BD}$ is an arbitrary constant. Using Eqs.~(\ref{eq2.3:5.4}), (\ref{eq2.6:5.7}) above, one finds that $- \, 2 \, \alpha_0 \, \varphi = \ln \Phi$ and that the (logarithmic) coupling function is simply
\begin{equation}
\label{eq2.8:5.9}
a(\varphi) = \alpha_0 \, \varphi + \text{const.} \, ,
\end{equation}
where $\alpha_0 = \mp (2\omega_\text{BD} + 3)^{-1/2}$, depending on the sign chosen in Eq.~(\ref{eq2.3:5.4}). Independently of this sign, one has the link
\begin{equation}
\label{eq2.9:5.10}
\alpha_0^2 = \frac{1}{2 \, \omega_\text{BD} + 3} \,.
\end{equation}
Note that $2 \, \omega_\text{BD} + 3$ must be positive for the spin-0 excitations to have the correct (non ghost) sign.

\smallskip

Let us now discuss the often considered case of a massive scalar field having a nonminimal coupling to curvature
\begin{equation}
\label{eq2.10:5.11}
S = \frac{c^4}{16 \pi \, G_*} \int \frac{d^4 x}{c} \ \tilde g^{1/2}
\left( \tilde R - \tilde g^{\mu\nu} \partial_{\mu} \Phi \,
\partial_{\nu} \Phi - m_{\Phi}^2 \Phi^2 + \xi \tilde R \, \Phi^2
\right) + S_m [\psi_m ; \tilde g_{\mu\nu}] \,.
\end{equation}
This is of the form (\ref{eq2.1:5.2}) with
\begin{equation}
\label{eq2.11:5.12}
F(\Phi) = 1 + \xi \Phi^2 \, , \ Z(\Phi) = 1 \, , \ U(\Phi) =
m_{\Phi}^2 \Phi^2 \,.
\end{equation}
The case $\xi = - \frac{1}{6}$ is usually referred to as that of `conformal coupling'. With the variables (\ref{eq2.10:5.11}) the theory is ghost-free only if $2 \, (1 + \xi \Phi^2)^2 \, (d\varphi / d\Phi)^2 = 1 + \xi (1 + 6 \, \xi) \, \Phi^2$ is everywhere positive. If we do not wish to restrict the initial values of $\Phi$, we must
have $\xi (1 + 6 \, \xi) > 0$. Introducing then the notation $\chi \equiv \sqrt{\xi (1+ 6 \, \xi)}$, we get the following link between $\Phi$ and $\varphi$:
\begin{eqnarray}
\label{eq2.12:5.13}
2 \sqrt 2 \, \varphi &= &\frac{\chi}{\xi} \, \ln \left[ 1+2 \, \chi
\, \Phi \left( \sqrt{1+\chi^2 \, \Phi^2} + \chi \, \Phi
\right)\right] \nonumber \\
&+ &\sqrt 6 \, \ln \left[ 1-2 \sqrt 6 \, \xi \, \Phi \, \frac{\sqrt{1
+ \chi^2 \, \Phi^2} - \sqrt 6 \, \xi \, \Phi}{1 + \xi \, \Phi^2}
\right] \,.
\end{eqnarray}
For small values of $\Phi$, this yields $\varphi = \Phi / \sqrt 2 + {\mathcal O} (\Phi^3)$. The potential and the coupling functions are given by
\begin{equation}
\label{eq2.13:5.14}
V(\varphi) = \frac{m_{\Phi}^2 \, \Phi^2}{1 + \xi \Phi^2} \, ,
\end{equation}
\begin{equation}
\label{eq2.14:5.15}
a(\varphi) = - \frac{1}{2} \, \ln (1 + \xi \Phi^2) \,.
\end{equation}

These functions have singularities when $1 + \xi \Phi^2$ vanishes. If we do not wish to restrict the initial value of $\Phi$ we must assume $\xi > 0$ (which then implies our previous assumption $\xi (1+ 6 \,
\xi) > 0$). Then there is a one-to-one relation between $\Phi$ and $\varphi$ over the entire real line. Small values of $\Phi$ correspond to small values of $\varphi$ and to a coupling function
\begin{equation}
\label{eq2.15:5.16}
a(\varphi) = - \, \xi \, \varphi^2 + {\mathcal O} (\varphi^4) \,.
\end{equation}
On the other hand, large values of $\vert \Phi \vert$ correspond to large values of $\vert \varphi \vert$, and to a coupling function of the asymptotic form
\begin{equation}
\label{eq2.16:5.17}
a(\varphi) \simeq - \sqrt 2 \ \frac{\xi}{\chi} \, \vert \varphi \vert + \text{const.}
\end{equation}
The potential $V(\varphi)$ has a minimum at $\varphi = 0$, as well as other minima at $\varphi \to \pm \infty$. If we assume, for instance, that $m_{\Phi}^2$ and the cosmological dynamics are such that the
cosmological value of $\varphi$ is currently attracted towards zero, the value of $\varphi$ at large distances from the local gravitating systems we shall consider will be $\varphi_0 \ll 1$.

\smallskip

As a final example of a possible tensor-scalar gravity theory, let us discuss the string-motivated dilaton-runaway scenario considered in \cite{Damour-Piazza-Veneziano02}. The starting action (a functional
of $\bar g_{\mu\nu}$ and $\Phi$) was taken of the general form
\begin{eqnarray}
S = \int d^4 x \, \sqrt{\bar g} \!\!\!\!\!\!\!\!\!\! &&\biggl( \frac{B_g (\Phi)}{\alpha'} \,
\bar R + \frac{B_{\Phi} (\Phi)}{\alpha'} \, [2 \, \bar\Box \, \Phi \nonumber \\
&& -(\bar\nabla \Phi)^2] - \frac{1}{4} \, B_F (\Phi) \bar F^2 - V (\Phi)
+ \cdots \biggl) \, , \nonumber
\end{eqnarray}
and it was assumed that all the functions $B_i (\Phi)$ have a regular asymptotic behavior when $\Phi \to +\infty$ of the form $B_i (\Phi) = C_i + {\mathcal O} (e^{-\Phi})$. Under this assumption the early
cosmological evolution can push $\Phi$ towards $+ \infty$ (hence the name `runaway dilaton'). In the canonical, `Einstein frame' representation (\ref{eq2.4:5.5}), one has, for large
values of $\Phi$, $\Phi \simeq c \, \varphi$, where $c$ is a numerical constant, and the coupling function to hadronic matter is given by
$$
e^{a(\varphi)} \propto \Lambda_{\rm QCD} (\varphi) \propto B_g^{-1/2}
(\varphi) \exp [-8\pi^2 \, b_3^{-1} \, B_F (\varphi)]
$$
where $b_3$ is the one-loop rational coefficient entering the renormalization-group running of the gauge field coupling $g_F^2$. This finally yields a coupling function of the approximate form (for
large values of $\varphi$):
$$
a(\varphi) \simeq k \, e^{-c \varphi} + \text{const.} \, ,
$$
where the dimensionless constants $k$ and $c$ are both expected to be of order unity. [The constant $c$ must be positive, but the sign of $k$ is not \textit{a priori} restricted.]

\smallskip

Summarizing: the JFBD model yields a coupling function which is a linear function of $\varphi$, Eq.~(\ref{eq2.8:5.9}), a nonminimally coupled scalar yields a coupling function which interpolates between a quadratic function of $\varphi$, Eq.~(\ref{eq2.15:5.16}), and a linear one, Eq.~(\ref{eq2.16:5.17}), and the dilaton-runaway scenario of Ref.~\cite{Damour-Piazza-Veneziano02} yields a coupling function of a decaying exponential type.

\subsection{The role of the coupling function $a(\varphi)$; definition of the two-dimensional space of tensor-scalar gravity theories $T_1 (\alpha_0 , \beta_0)$}\label{ssec:5.4}

Let us now discuss how the coupling function $a(\varphi)$ enters the observable predictions of tensor-scalar gravity at the first post-Newtonian (1PN) level, i.e., in the weak-field conditions appropriate to
solar-system tests. It was shown in previous work that, if one uses appropriate units in the asymptotic region far from the system, namely units such that the asymptotic value $a (\varphi_0)$ of $a(\varphi)$ vanishes\footnote{In these units the Einstein metric $g_{\mu\nu}^*$ and the physical metric $\tilde g_{\mu\nu}$ asymptotically coincide.}, all observable quantities at the 1PN level depend only on the values of the first two derivatives of the $a(\varphi)$ at $\varphi = \varphi_0$. More precisely, if one defines
\begin{equation}
\label{eq2.17:5.18}
\alpha (\varphi) \equiv \frac{\partial \, a(\varphi)}{\partial \,
\varphi} \, ; \ \beta (\varphi) \equiv \frac{\partial \, \alpha
(\varphi)}{\partial \, \varphi} = \frac{\partial^2 \,
a(\varphi)}{\partial \, \varphi^2} \, ,
\end{equation}
and denotes by $\alpha_0 \equiv \alpha (\varphi_0)$, $\beta_0 \equiv \beta (\varphi_0)$ their asymptotic values, one finds (see, e.g., \cite{DEF92}) that the effective gravitational constant between two bodies (as measured by a Cavendish experiment) is given by
\begin{equation}
\label{eq2.18:5.19}
G = G_* (1 + \alpha_0^2) \, ,
\end{equation}
while, among the PPN parameters, only the two basic Eddington ones, $\bar\gamma \equiv \gamma^{\rm PPN}-1$, and $\bar\beta \equiv \beta^{\rm PPN}-1$, do not vanish, and are given by
\begin{eqnarray}
\label{eq2.19a:5.20a}
\bar\gamma \equiv \gamma^{\rm PPN} - 1 &= & - 2 \, \frac{\alpha_0^2}{1 + \alpha_0^2}
\, , \\
\label{eq2.19b:5.20b}
\bar\beta \equiv \beta^{\rm PPN} - 1 &= & \frac{1}{2} \ \frac{\alpha_0 \, \beta_0 \,
\alpha_0}{(1 + \alpha_0^2)^2} \,.
\end{eqnarray}
The structure of the results (\ref{eq2.19a:5.20a}) and (\ref{eq2.19b:5.20b}) can be transparently expressed by means of simple (Feynman-like) diagrams (see, e.g., \cite{DEF96a}). Eqs.~(\ref{eq2.18:5.19}) and (\ref{eq2.19a:5.20a}) correspond to diagrams where the interaction between two worldlines (representing two massive bodies) is mediated by the sum of the exchange of one graviton and one scalar particle. The scalar couples to matter with strength $\sim \alpha_0 \, \sqrt{G_*}$. The exchange of a scalar excitation then leads to a term $\propto \alpha_0^2$. On the other hand, Eq.~(\ref{eq2.19b:5.20b}) corresponds to a nonlinear interaction between three worldlines involving: (i) the `generation' of a scalar excitation on a first worldline (factor $\alpha_0$), (ii) a nonlinear vertex on a second worldline associated to the quadratic piece of
$a(\varphi)$ ($a_{\rm quad} (\varphi) = \frac{1}{2} \, \beta_0 (\varphi - \varphi_0)^2$; so that one gets a factor $\beta_0$), and (iii) the final `absorption' of a scalar excitation on a third worldline
(second factor $\alpha_0$).

\smallskip

Eqs.~(\ref{eq2.19a:5.20a}) and (\ref{eq2.19b:5.20b}) can be summarized by saying that the first two coefficients in the Taylor expansion of the coupling function $a(\varphi)$ around $\varphi = \varphi_0$ (after setting $a(\varphi_0) = 0$)
\begin{equation}
\label{eq2.20:5.21}
a(\varphi) = \alpha_0 (\varphi - \varphi_0) + \frac{1}{2} \, \beta_0
(\varphi - \varphi_0)^2 + \cdots
\end{equation}
suffice to determine the quasi-stationary, weak-field (1PN) predictions of any tensor-scalar theory. In other words, the solar-system tests only explore the `osculating approximation' (\ref{eq2.20:5.21}) (slope and local curvature) to the function $a(\varphi)$. Note that GR corresponds to a vanishing coupling
function $a(\varphi) = 0$ (so that $\alpha_0 = \beta_0 = \cdots = 0$), the JFBD model corresponds to keeping only the first term on the R.H.S. of (\ref{eq2.20:5.21}), while, for instance, the nonminimally coupled scalar field (with asymptotic value $\varphi_0 \ll 1$) does indeed lead to nonzero values for both $\alpha_0$ and $\beta_0$, namely
\begin{equation}
\label{eq2.21:5.22}
\alpha_0 \simeq - \, 2 \, \xi \, \varphi_0 \, ; \ \beta_0 \simeq - \,
2 \, \xi \,.
\end{equation}

Finally the dilaton-runaway scenario considered above leads also to non zero values for both $\alpha_0$ and $\beta_0$, namely
\begin{equation}
\label{eq2.22:5.23}
\alpha_0 \simeq - \, k \, c \, e^{-c\varphi_0} \, ; \ \beta_0 \simeq
+ \, k \, c^2 \, e^{-c\varphi_0} \, ,
\end{equation}
for a largish value of $\varphi_0$. Note that the dilaton-runaway model naturally predicts that $\alpha_0 \ll 1$, and that $\beta_0$ is of the same order of magnitude as $\alpha_0 : \beta_0 \simeq - \, c \, \alpha_0$ with $c$ being (positive and) of order unity. The interesting outcome is that such a model is well approximated by the usual JFBD model (with $\beta_0 = 0$). This shows that a JFBD-like theory could come out from a model which is initially quite different from the usual exact JFBD theory.

\smallskip

As we shall discuss in detail below, solar-system tests constrain $\alpha_0^2$ and $\alpha_0^2 \, \vert \beta_0 \vert$ to be both
small. This immediately implies that $|\alpha_0|$ must be small, i.e., that the scalar field is linearly weakly coupled to matter. On the other hand, the quadratic coupling parameter $\beta_0$ is not
directly constrained. Both its magnitude and its sign can be more or less arbitrary. Note that there are no \textit{a priori} sign restrictions on $\beta_0$. The conformal factor $A^2 (\varphi) = \exp (2 \, a(\varphi))$ entering Eq.~(\ref{eq2.4:5.5}) had to be positive, but this leads to no restrictions on the sign of $a(\varphi)$ and of its various derivatives\footnote{As explained above, we assume here the presence of a potential term $V(\varphi)$ to fix the asymptotic value $\varphi_0$ of $\varphi$. If the potential $V(\varphi)$ is absent (or negligible), the `attractor mechanism' of Refs.~\cite{DN93,Damour-Polyakov94} would attract $\varphi$ to a {\it minimum} of the coupling function $a(\varphi)$, thereby favoring a {\it positive} value of $\beta_0$.}. For instance, in the nonminimally coupled scalar field case, it seemed more natural to require $\xi > 0$, which leads to a negative $\beta_0$ in view of Eq.~(\ref{eq2.21:5.22}).

\smallskip

Let us summarize the results above: (i) the most general tensor-scalar theory\footnote{Under the assumption that the potential $V(\varphi)$ is a slowly-varying function of $\varphi$, which modifies the propagation of $\varphi$ only on very large scales.} is described by one arbitrary function $a(\varphi)$; and (ii) weak-field tests depend only on the first two terms, parametrized by $\alpha_0$ and $\beta_0$, in the Taylor expansion (\ref{eq2.20:5.21}) of $a(\varphi)$ around its asymptotic value $\varphi_0$. 

\smallskip

From this follows a rather natural way to define a simple {\it mini space of tensor-scalar theories}. It suffices to consider the two-dimensional space of theories, say $T_1 (\alpha_0 , \beta_0)$, defined by the coupling
function which is a quadratic polynomial in $\varphi$ \cite{DEF93,DEF96}, say
\begin{equation}
\label{eq2.23:5.24}
a_{\alpha_0 , \beta_0} (\varphi) = \alpha_0 (\varphi - \varphi_0) +
\frac{1}{2} \, \beta_0 (\varphi - \varphi_0)^2 \,.
\end{equation}
As indicated, this class of theories depends only on two parameters: $\alpha_0$ and $\beta_0$. The asymptotic value $\varphi_0$ of $\varphi$ does not count as a third parameter (when using the form
(\ref{eq2.23:5.24})) because one can always work with the shifted field $\bar\varphi \equiv \varphi - \varphi_0$, with asymptotic value $\bar\varphi_0 = 0$ and coupling function $a_{\alpha_0 , \beta_0}
(\bar\varphi) = \alpha_0 \, \bar\varphi + \frac{1}{2} \, \beta_0 \, \bar\varphi^2$. Moreover, as already said, the asymptotic value $a(\varphi_0)$ of $a(\varphi)$ has also no physical meaning, because
one can always use units such that it vanishes (as done in (\ref{eq2.23:5.24})). 

\smallskip

Note also that an alternative way to represent the same class of theories is to use a coupling function of the very
simple form
\begin{equation}
\label{eq2.24:5.25}
a_{\beta} (\varphi) = \frac{1}{2} \, \beta \, \varphi^2 \, ,
\end{equation}
but to keep the asymptotic value $\varphi_0$ as an independent parameter. This class of theories is clearly equivalent to $T_1 (\alpha_0 , \beta_0)$, Eq.~(\ref{eq2.23:5.24}), with the dictionary: $\alpha_0 = \beta \, \varphi_0$, $\beta_0 = \beta$.

\subsection{Tensor-scalar gravity, strong-field effects, and binary-pulsar observables}\label{ssec:5.5}

Having chosen some mini-space of gravity theories, we now wish to derive what predictions these theories make for the timing observables of binary pulsars. To do this we need to generalize the general relativistic treatment of the motion and timing of binary systems comprising strongly self-gravitating bodies summarized above. Let us recall that this treatment was based on a multi-chart method, using a {\it matching} between two separate problems: (i) the `internal problem' considers each strongly self-gravitating body in a suitable approximately freely falling frame where the influence of its companion is small, and (ii) the `external problem' where the two bodies are described as effective point masses which interact via the various fields they are coupled to. Let us first consider the {\it internal problem}, i.e., the description of a neutron star in an approximately freely falling frame where the influence of the companion is reduced to imposing some boundary conditions on the tensor and scalar fields with which it interacts
\cite{Eardley75,WillEardley77,DEF92,DEF93,DEF96}. The field equations of a general tensor-scalar theory, as derived from the canonical action (\ref{eq2.4:5.5}) (neglecting the effect of $V(\varphi)$) read
\begin{eqnarray}
\label{eq3.1a:5.26a}
R_{\mu\nu}^* &= & 2 \, \partial_{\mu} \varphi \, \partial_{\nu}
\varphi + 8 \pi \, G_* \left(T_{\mu\nu}^* - \frac{1}{2} \, T^*
g_{\mu\nu}^* \right) \, , \\
\label{eq3.1b:5.26b}
\Box_{g_*} \, \varphi &= & - \, 4\pi \, G_* \, \alpha (\varphi) \,
T_* \, ,
\end{eqnarray}
where $T_*^{\mu\nu} \equiv 2 \, c \, (g_*)^{-1/2} \, \delta S_m / \delta g_{\mu\nu}^*$ denotes the material stress-energy tensor in `Einstein units', and $\alpha (\varphi)$ the $\varphi$-derivative of the coupling function, see Eq.~(\ref{eq2.17:5.18}). All tensorial operations in Eqs.~(\ref{eq3.1a:5.26a}) and (\ref{eq3.1b:5.26b}) are performed by using the Einstein metric $g_{\mu\nu}^*$.

\smallskip

Explicitly writing the field equations (\ref{eq3.1a:5.26a}) and (\ref{eq3.1b:5.26b}) for a slowly rotating (stationary, axisymmetric) neutron star, labelled\footnote{We henceforth use the labels $A$ and $B$ for the (recycled) pulsar and its companion, instead of the labels $a$ and $b$ used above. We henceforth use the label $a$ to denote the {\it asymptotic} value of some quantity (at large radial distances within the local frame, $X_A^i$ or $X_B^i$, of the considered neutron star $A$ or $B$).} $A$, leads to a coupled set of ordinary differential equations constraining the radial dependence of $g_{\mu\nu}^*$ and $\varphi$ \cite{DEF96,Hartle67}. Imposing the boundary conditions $g_{\mu\nu}^* \to \eta_{\mu\nu}$, $\varphi \to \varphi_a$ at large radial distances, finally determines the crucial `form factors' (in Einstein units) describing the effective coupling between the neutron star $A$ and the fields to which it is sensitive: total mass $m_A (\varphi_a)$,
total scalar charge $\omega_A (\varphi_a)$, and inertia moment $I_A (\varphi_a)$. As indicated, these quantities are functions of the {\it asymptotic} value $\varphi_a$ of $\varphi$ felt by the considered neutron
star\footnote{This $\varphi_a$ is a combination of the cosmological background value $\varphi_0$ and of the scalar influence of the companion of the considered neutron star. It varies with the orbital period and
is determined as part of the `external problem' discussed below. Note that, strictly speaking, the label $a$ (for {\it asymptotic}) should be indexed by the label of the considered neutron star: i.e. one should use a label $a_A$ (and a {\it locally asymptotic} value $\varphi_{a_A}$) when considering the neutron star $A$, and a label $a_B$ (with a corresponding $\varphi_{a_B}$) when considering the neutron star $B$.}. They satisfy the relation $\omega_A (\varphi_a) = - \partial \, m_A (\varphi_a) / \partial \, \varphi_a$. From them, one defines other
quantities that play an important role in binary pulsar physics, notably
\begin{equation}
\label{eq3.2:5.27}
\alpha_A (\varphi_a) \equiv - \frac{\omega_A}{m_A} \equiv
\frac{\partial \ln m_A}{\partial \, \varphi_a} \, ,
\end{equation}
\begin{equation}
\label{eq3.3:5.28}
\beta_A (\varphi_a) \equiv \frac{\partial \, \alpha_A}{\partial \,
\varphi_a} \, ,
\end{equation}
as well as
\begin{equation}
\label{eq3.4:5.29}
k_A (\varphi_a) \equiv - \frac{\partial \ln I_A}{\partial \,
\varphi_a} \,.
\end{equation}
The quantity $\alpha_A$, Eq.~(\ref{eq3.2:5.27}), plays a crucial role. It measures the {\it effective coupling strength} between the neutron star and the ambient scalar field. If we formally let the
self-gravity of the neutron $A$ tend toward zero (i.e., if we consider a weakly self-gravitating object), the function $\alpha_A (\varphi_a)$ becomes replaced by $\alpha (\varphi_a)$ where $\alpha (\varphi) \equiv \partial \, a (\varphi) / \partial \, \varphi$ is the coupling strength appearing in the R.H.S. of Eq.~(\ref{eq3.1b:5.26b}). Roughly speaking, we can think of $\alpha_A (\varphi_a)$ as a (suitable defined) average value of the local coupling strength $\alpha (\varphi (r))$ over the radial profile of the neutron star $A$.

%
%
\begin{figure}[h]
\centering
\includegraphics[height=5cm]{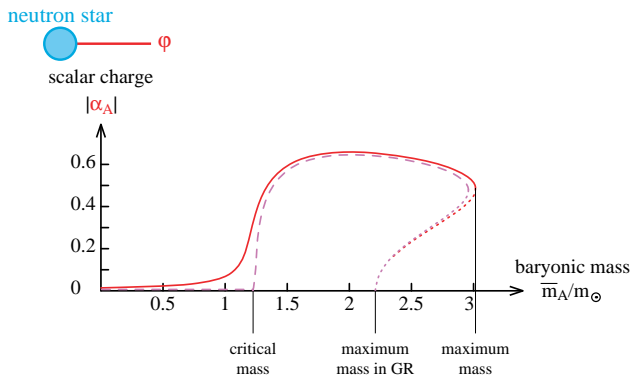}
%
%
\caption{Dependence upon the baryonic mass $\bar m_A$ of the coupling parameter $\alpha_A$ in the theory $T_1 (\alpha_0 , \beta_0)$ with $\alpha_0 = -0.014$, $\beta_0 = -6$. Figure taken from \cite{Esposito-Farese05}.}
\label{fig:2}       
\end{figure}

It was pointed out in Refs.~\cite{DEF93,DEF96} that the strong self-gravity of a neutron star can cause the effective coupling strength $\alpha_A (\varphi_a)$ to become of order unity, even when its weak-field counterpart $\alpha_0 = \alpha (\varphi_a)$ is extremely small (as is implied by solar-system tests that put strong constraints on the PPN combination $\bar\gamma = - 2 \, \alpha_0^2 / (1+\alpha_0^2)$). This is illustrated, in the minimal context of the $T_1 (\alpha_0 , \beta_0)$ class of theories, in Figure~\ref{fig:2}.

\smallskip

Note that when the baryonic mass $\bar m_A$ of the neutron star is smaller than the critical mass $\bar m_{cr} \simeq 1.24 \, M_{\odot}$ the effective scalar coupling strength $\alpha_A$ of the star is quite small (because it is proportional to its weak-field limit $\alpha_0 = \alpha (\varphi_a)$). By contrast, when $\bar m_A > \bar m_{cr}$, $\vert \alpha_A \vert$ becomes of order unity, nearly independently of the externally imposed $\alpha_0 = \alpha_a = \alpha (\varphi_a)$. This interesting {\it non-perturbative} behaviour was related in \cite{DEF93,DEF96} to a mechanism of {\it spontaneous scalarization}, akin to the well-known mechanism of spontaneous magnetization of ferromagnets. See also \cite{DEF06} for a simple analytical description of the behaviour of $\alpha_A$.

\smallskip

Let us also mention in passing that, in the case where $A$ is a black hole, the effective coupling strength $\alpha_A$ actually {\it vanishes} \cite{DEF92}. This result is related to the impossibility of having (regular) `scalar hair' on a black hole.

\smallskip

We have sketched above the first part of the matching approach to the motion and timing of strongly self-gravitating bodies: the `internal problem'. It remains to describe the remaining `external problem'. As already mentionned (and emphasized, in the present context, by Eardley \cite{Eardley75,Will93}), the most efficient way to describe the external problem is, instead of matching in detail the external
fields $(g_{\mu\nu}^* , \varphi)$ to the fields generated by each body in its comoving frame, to `skeletonize' the bodies by point masses. Technically this means working with the action
\begin{eqnarray}
\label{eq3.8:5.30}
S &= &\frac{c^4}{16\pi \, G_*} \int \frac{d^D x}{c} \, g_*^{1/2} [R_* -
2 \, g_*^{\mu\nu} \, \partial_{\mu} \varphi \, \partial_{\nu}
\varphi] \nonumber \\
&- &\sum_A c \int m_A (\varphi (z_A)) (- g_{\mu\nu}^* (z_A) \,
dz_A^{\mu} \, dz_A^{\nu})^{1/2} \, ,
\end{eqnarray}
where the function $m_A (\varphi)$ in the last term on the R.H.S. is the function $m_A (\varphi_a)$ obtained above by solving the internal problem. Eq.~(\ref{eq3.8:5.30}) indicates that the argument of this
function is taken to be $\varphi_a = \varphi (z_A)$, i.e., the value that the scalar field (as viewed in the external problem) takes at the location $z_A^{\mu}$ of the center of mass of body $A$. However, as body $A$ is described, in the external problem, as a point mass this causes a technical difficulty: the externally determined field $\varphi (x)$ becomes formally singular at the location of the point sources, so that $\varphi (z_A)$ is \textit{a priori} undefined. One can either deal with this problem by coming back to the physically well-defined matching approach (which shows that $\varphi (z_A)$ should be replaced by $\varphi_a$, the value of $\varphi$ in an intermediate domain $R_A \ll r \ll \vert \mbox{\boldmath$z$}_A - \mbox{\boldmath$z$}_B \vert$), or use the efficient technique of dimensional regularization. This means that the spacetime dimension $D$ in Eq.~(\ref{eq3.8:5.30}) is first taken to have a complex value such that $\varphi (z_A)$ is finite, before being analytically continued
to its physical value $D=4$.

\smallskip

One then derives from the action (\ref{eq3.8:5.30}) two important consequences for the motion and timing of binary pulsars. First, one derives the Lagrangian describing the relativistic interaction
between $N$ strongly self-gravitating bodies (including orbital $\sim (v/c)^2$ effects, and neglecting ${\mathcal O} (v^4 / c^4)$ ones) \cite{Will93,Eardley75,DEF92,WillZaglauer89}. It is the sum of
one-body, two-body and three-body terms.

\smallskip

The one-body action has the usual form of the sum (over the label $A$) of the kinetic term of each point mass:
\begin{eqnarray}
\label{eq3.9:5.31}
L_A^{\mbox{\scriptsize one-body}} &= &- m_A \, c^2 \sqrt{1 -
\mbox{\boldmath$v$}_A^2 / c^2} \nonumber \\
&= &-m_A \, c^2 + \frac{1}{2} \, m_A \,
\mbox{\boldmath$v$}_A^2 + \frac{1}{8} \, m_A
\frac{(\mbox{\boldmath$v$}_A^2)^2}{c^2} + {\mathcal O} \left(
\frac{1}{c^4} \right) \,.
\end{eqnarray}
Here, we use Einstein units, and the inertial mass $m_A$ entering Eq.~(\ref{eq3.9:5.31}) is $m_A \equiv m_A (\varphi_0)$, where $\varphi_0$ is the asymptotic value of $\varphi$ far away from the considered
$N$-body system.

\smallskip

The two-body action is a sum over the pairs $A,B$ of a term $L_{AB}^{\mbox{\scriptsize 2-body}}$ which differs from the GR-predicted 2-body Lagrangian in two ways: (i) the usual gravitational constant $G$ appearing as an overall factor in $L_{AB}^{\mbox{\scriptsize 2-body}}$ must be replaced by an effective
(body-dependent) gravitational constant (in the appropriate units mentioned above) given by
\begin{equation}
\label{eq3.10:5.32}
G_{AB} = G_* (1 + \alpha_A \, \alpha_B) \, ,
\end{equation}
and (ii) the relativistic $({\mathcal O} (v^2/c^2))$ terms in $L_{AB}^{\mbox{\scriptsize 2-body}}$ contain, in addition to those predicted by GR, new velocity-dependent terms of the form
\begin{equation}
\label{eq3.11:5.33}
\delta^{\gamma} L_{AB}^{\mbox{\scriptsize 2-body}} = (\bar\gamma_{AB}) \frac{G_{AB} \, m_A \, m_B}{r_{AB}} \frac{(\mbox{\boldmath$v$}_A -
\mbox{\boldmath$v$}_B)^2}{c^2} \, ,
\end{equation}
with
\begin{equation}
\label{eq3.12:5.34}
\bar\gamma_{AB} \equiv \gamma_{AB} - 1 = - \, 2 \, \frac{\alpha_A \, \alpha_B}{1 + \alpha_A \, \alpha_B} \,.
\end{equation}
In these expressions $\alpha_A \equiv \alpha_A (\varphi_0) \equiv \partial \ln m_A (\varphi_0) / \partial \varphi_0$ (see Eq.~(\ref{eq3.2:5.27}) with $\varphi_a \to \varphi_0$).

\smallskip

Finally, the 3-body action is a sum over the pairs $B,C$ and over $A$ (with $A \ne B$, $A \ne C$, but the possibility of having $B=C$) of
\begin{equation}
\label{eq3.13:5.35}
L_{ABC}^{\mbox{\scriptsize 3-body}} = - (1+2 \, \bar\beta_{BC}^A ) \,
\frac{G_{AB} \, G_{AC} \, m_A \, m_B \, m_C}{c^2 \, r_{AB} \, r_{AC}}
\end{equation}
where
\begin{equation}
\label{eq3.14:5.36}
\bar\beta_{BC}^A \equiv \beta_{BC}^A - 1 = \frac{1}{2} \, \frac{\alpha_B \, \beta_A \,
\alpha_C}{(1+ \alpha_A \, \alpha_B) (1 + \alpha_A \, \alpha_C)} \, ,
\end{equation}
with $\beta_A = \partial \alpha_A (\varphi_0) / \partial \varphi_0$ (see Eq.~(\ref{eq3.3:5.28}) with $\varphi_a \to \varphi_0$).

\smallskip

When comparing the strong-field results (\ref{eq3.10:5.32}), (\ref{eq3.12:5.34}), (\ref{eq3.14:5.36}) to their weak-field counterparts (\ref{eq2.18:5.19}), (\ref{eq2.19a:5.20a}), (\ref{eq2.19b:5.20b}) one sees that the body-dependent quantity $\alpha_A$ replaces the weak-field coupling strength
$\alpha_0$ in all quantities which are linked to a scalar effect generated by body $A$. Note also that, in keeping with the `3-body' nature of Eq.~(\ref{eq3.13:5.35}), the quantity $\beta_{BC}^A - 1$ is linked to scalar interactions which are generated in bodies $B$ and $C$ and which nonlinearly interact on body $A$. The notation used above has been chosen to emphasize that $\gamma_{AB}$ and $\beta_{BC}^A$ are strong-field analogs of the usual Eddington parameters $\gamma^{\rm PPN}$, $\beta^{\rm PPN}$, so that $\bar\gamma_{AB}$ and $\bar\beta_{BC}^A$ are strong-field analogs of the `post-Einstein' 1PN parameters $\bar\gamma$ and $\bar\beta$ (which vanish in GR). Indeed the usual PPN results for the post-Einstein terms in the ${\mathcal O} (1/c^2)$ $2$-body and $3$-body Lagrangians are obtained by replacing in Eqs.~(\ref{eq3.11:5.33}) and (\ref{eq3.13:5.35}) $\bar\gamma_{AB} \to \bar\gamma$, $\bar\beta_{BC}^A \to \bar\beta$ and $G_{AB} \to G$.

\smallskip

The non-perturbative strong-field effects discussed above show that the strong self-gravity of neutron stars can cause $\gamma_{AB}$ and $\beta_{BC}^A$ to be significantly different from their GR values $\gamma^{\rm GR} = 1$, $\beta^{\rm GR} = 1$, in some scalar-tensor theories having a small value of the basic coupling parameter $\alpha_0$ (so that $\gamma^{\rm PPN} - 1 \propto \alpha_0^2$ and
$\beta^{\rm PPN} - 1 \propto \beta_0 \, \alpha_0^2$ are both small). For instance, Fig.~\ref{fig:2} shows that it is possible to have $\alpha_A \sim \alpha_B \sim \pm \, 0.6$ which implies $\gamma_{AB} - 1 \sim -
\, 0.53$, i.e., a 50\% deviation from GR! Even larger effects can arise in $\beta_{BC}^A - 1$ because of the large values that $\beta_A = \partial \alpha_A / \partial \varphi_0$ can reach near the spontaneous scalarization transition \cite{DEF96}.

\smallskip

Those possible strong-field modifications of the effective Eddington parameters $\gamma_{AB}$, $\beta_{BC}^A$, which parametrize the `first post-Keplerian' (1PK) effects (i.e., the orbital effects
$\sim v^2 / c^2$ smaller than those entailed by the Lagrangian $\underset{A}{\sum} \frac{1}{2} \, m_A \, \mbox{\boldmath$v$}_A^2 + \frac{1}{2} \underset{A \ne B}{\sum} G_{AB} \, m_A \, m_B / r_{AB}$),
can then significantly modify the usual GR predictions relating the directly observable parametrized post-Keplerian (PPK) parameters to the values of the masses of the pulsar and its companion. As worked out in Refs.~\cite{Will93,Damour-Taylor92,DEF92,DEF96} one finds the following modified predictions for the PPK parameters $k \equiv \langle \dot\omega \rangle / n$, $r$ and $s$:
\begin{eqnarray}
\label{eq3.15:5.37}
k^{\rm th} (m_A , m_B) &= &\frac{3}{1-e^2} \left( \frac{G_{AB} (m_A +
m_B) \, n}{c^3} \right)^{2/3} \nonumber \\
&&\left[ \frac{1 - \frac{1}{3} \, \alpha_A \, \alpha_B}{1 + \alpha_A
\, \alpha_B} - \frac{X_A \, \beta_B \, \alpha_A^2 + X_B \, \beta_A \,
\alpha_B^2}{6 \, (1 + \alpha_A \, \alpha_B)^2} \right] \, ,
\end{eqnarray}
\begin{equation}
\label{eq3.16:5.38}
r^{\rm th} (m_A , m_B) = G_{0B} \, m_B \, ,
\end{equation}
\begin{equation}
\label{eq3.17:5.39}
s^{\rm th} (m_A , m_B) = \frac{n \, x_A}{X_B} \left[ \frac{G_{AB}
(m_A + m_B) \, n}{c^3} \right]^{-1/3} \,.
\end{equation}
Here, the label $A$ refers to the object which is timed (`the pulsar'\footnote{In the double binary pulsar, both the first discovered pulsar and its companion are pulsars. However, the companion $B$ is a non recycled, slow pulsar whose motion is well described by Keplerian parameters only.}), the label $B$ refers to its companion, $x_A = a_A \sin i/c$ denotes the projected semi-major axis of the orbit of $A$ (in light seconds), $X_A \equiv m_A / (m_A + m_B)$ and $X_B \equiv m_B / (m_A + m_B) = 1 - X_A$ the mass ratios, $n \equiv 2\pi / P_b$ the orbital frequency and $G_{0B} = G_* (1 + \alpha_0 \, \alpha_B)$ the effective gravitational constant measuring the interaction between $B$ and a test object (namely electromagnetic waves on their way from the pulsar toward the Earth). In addition one must replace the unknown bare Newtonian $G_*$ by its expression in terms of the one measured in Cavendish experiments, i.e., $G_* = G / (1 + \alpha_0^2)$ as deduced from Eq.~(\ref{eq2.18:5.19}).

\smallskip

The modified theoretical prediction for the PPK parameter $\gamma$ entering the `Einstein time delay' $\Delta_E$, Eq.~(\ref{eq3.16}), is more complicated to derive because one must take into account the
modulation of the proper spin period of the pulsar caused by the variation of its moment of inertia $I_A$ under the (scalar) influence of its companion \cite{Will93,Eardley75,DEF96}. This leads to
\begin{eqnarray}
\label{eq3.18:5.40}
\gamma^{\rm th} (m_A , m_B) &= &\frac{e}{n} \, \frac{X_B}{1 + \alpha_A
\, \alpha_B} \left( \frac{G_{AB} (m_A + m_B) \, n}{c^3} \right)^{2/3} \nonumber \\
&&[X_B (1 + \alpha_A \, \alpha_B) + 1 + k_A \, \alpha_B ] \, ,
\end{eqnarray}
where $k_A (\varphi_0) = - \partial \ln I_A (\varphi_0) / \partial \varphi_0$ (see Eq.~(\ref{eq3.4:5.29}) with $\varphi_a \to \varphi_0$). Numerical studies \cite{DEF96} show that $k_A$ can take quite large
values. Actually, the quantity $k_A \, \alpha_B$ entering (\ref{eq3.18:5.40}) blows up near the scalarization transition when $\alpha_0 \to 0$ (keeping $\beta_0 < 0$ fixed). In other words a
theory which is closer to GR in weak-field conditions predicts larger deviations in the strong-field regime.

\smallskip

The structure dependence of the effective gravitational constant $G_{AB}$, Eq.~(\ref{eq3.10:5.32}), has also the consequence that the object $A$ does not fall in the same way as $B$ in the gravitational
field of the Galaxy. As most of the mass of the Galaxy is made of non strongly-self-gravitating bodies, $A$ will fall toward the Galaxy with an acceleration $\propto G_{A0}$, while $B$ will fall with an
acceleration $\propto G_{B0}$. Here, as above, $G_{A0} = G_{0A} = G_* (1 + \alpha_0 \, \alpha_A)$ is the effective gravitational constant between $A$ and any weakly self-gravitating body. As pointed out in Ref.~\cite{DS91} this possible violation of the universality of free fall of self-gravitating bodies can be constrained by using observational data on the class of small-eccentricity long-orbital-period binary pulsars. More precisely, the quantity which can be observationally constrained is not exactly the violation $\Delta_{AB} = (G_{0A} - G_{0B}) / G
=(1 + \alpha_0^2)^{-1} (\alpha_0 \, \alpha_A - \alpha_0 \, \alpha_B)
$ of the strong equivalence principle [which simplifies to $\Delta_{A0} = (G_{0A} - G) / G = (1 + \alpha_0^2)^{-1} (\alpha_0 \, \alpha_A - \alpha_0^2)$ in the case of observational relevance where one neglects the self-gravity of the white-dwarf companion] but rather\footnote{This refinement is given here for pedagogical completeness. However, in practice, the lowest-order result $\Delta \simeq (1 + \alpha_0^2)^{-1} (\alpha_0 \, \alpha_A - \alpha_0^2) \simeq \alpha_0 \, \alpha_A - \alpha_0^2$ is accurate enough.}
\cite{DEF92}
\begin{eqnarray}
\label{eq3.19:5.41}
\Delta^{\rm effective} &\equiv &\left( \frac{2 \, \gamma_{AB} - (X_A \,
\beta_{AA}^B + X_B \, \beta_{BB}^A) + 2}{3} \right)^{-1} \nonumber \\
&&(1 + \alpha_A \, \alpha_B)^{-3/2}
(1 + \alpha_0^2)^{-1} (\alpha_0 \, \alpha_A - \alpha_0 \, \alpha_B)
 \,.
\end{eqnarray}
Here, the index $B$ ($=$ white-dwarf companion) can be replaced by $0$ (weakly self-gravitating body) so that, for instance, $\gamma_{AB} = \gamma_{A0} = 1 - 2 \, \alpha_A \, \alpha_0 / (1 + \alpha_A \,
\alpha_0) = (1 - \alpha_A \, \alpha_0) / (1 + \alpha_A \, \alpha_0)$, as deduced from Eq.~(\ref{eq3.12:5.34}).

\smallskip

It remains to discuss the possible strong-field modifications of the theoretical prediction for the orbital period derivative $\dot P_b = \dot P_b^{\rm th} (m_A , m_B)$. This is obtained by deriving from the
effective action (\ref{eq3.8:5.30}) the energy lost by the binary system in the form of fluxes of spin-2 and spin-0 waves at infinity. The needed results in a generic tensor-scalar theory were derived in
Refs.~\cite{DEF92,WillZaglauer89} (in addition one must take into account the tensor-scalar modification of the additional `varying-Doppler' contribution to the observed $\dot P_b$ due to the Galactic acceleration \cite{DamourTaylor91}). The final result for $\dot P_b$ is of the form
\begin{eqnarray}
\label{eq3.20:5.42}
\dot P_b^{\rm th} (m_A , m_B) &= &\dot P_{b\varphi}^{\rm monopole} +
\dot P_{b\varphi}^{\rm dipole} + \dot P_{b\varphi}^{\rm quadrupole} +
\dot P_{bg^*}^{\rm quadrupole} \nonumber \\
&+ &\dot P_{b \, {\rm GR}}^{\rm
galagtic} + \delta^{\rm th} \, \dot P_{b}^{\rm galactic} \, ,
\end{eqnarray}
where, for instance, $\dot P_{b\varphi}^{\rm monopole}$ is (heuristically\footnote{Contrary to the GR case where a lot of effort was spent to show how the observed $\dot P_b$ was directly related to
the GR predictions for the $(v/c)^5$-accurate orbital equations of motion of a binary system \cite{Damour83}, we use here the indirect and less rigorous argument that the energy flux at infinity should be balanced by a corresponding decrease of the mechanical energy of the binary system.}) related to the monopolar flux of spin-0 waves at infinity. The term $\dot P_{bg^*}^{\rm quadrupole}$ corresponds to the usual quadrupolar flux of spin-2 waves at infinity. It reads:
\begin{eqnarray}
\label{eq3.20bis:5.43}
\dot P_{bg^*}^{\rm quadrupole} (m_A , m_B) &= &- \frac{192 \pi}{5 (1
+ \alpha_A \, \alpha_B)} \, \frac{m_A \, m_B}{(m_A + m_B)^2} \\
&& \left( \frac{G_{AB} (m_A + m_B) \, n}{c^3} \right)^{5/3} \frac{1 +
73 \, e^2 / 24 + 37 \, e^4 / 96}{(1-e^2)^{7/2}} \, , \nonumber
\end{eqnarray}
with $G_{AB} = G_* (1 + \alpha_A \, \alpha_B) = G (1 + \alpha_A \, \alpha_B) / (1 + \alpha_0^2)$, where $G_*$ is the `bare' gravitational constant appearing in the action, while $G$ is the gravitational constant measured in Cavendish experiments. The flux (\ref{eq3.20bis:5.43}) is the only one which survives in GR (although without any $\alpha_A$-related modifications). Among the several other contributions which arise in tensor-scalar theories, let us only write down the explicit expression of the contribution to
(\ref{eq3.20:5.42}) coming from the {\it dipolar flux of scalar waves}. Indeed, this contribution is, in most cases, the dominant one \cite{Eardley75} because it scales as $(v/c)^3$, while the monopolar and quadrupolar contributions scale as $(v/c)^5$. It reads
\begin{equation}
\label{eq3.21:5.44}
\dot P_{b\varphi}^{\rm dipole} (m_A , m_B) = -2\pi \, \frac{G_* \,
m_A \, m_B \, n}{c^3 (m_A + m_B)} \, \frac{1 + e^2 /
2}{(1-e^2)^{5/2}} \, (\alpha_A - \alpha_B)^2 \,.
\end{equation}
Note that the dipolar effect (\ref{eq3.21:5.44}) vanishes when $\alpha_A = \alpha_B$. Indeed, a binary system made of two identical objects $(A=B)$ cannot select a preferred direction for a dipole vector, and
cannot therefore emit any dipolar radiation. This also implies that double neutron star systems (which tend to have $m_A \approx m_B \sim 1.35 \, M_{\odot}$) will be rather poor emitters of dipolar radiation
(though (\ref{eq3.21:5.44}) still tends to dominate over the other terms in (\ref{eq3.20:5.42}), because of the remaining difference $(m_A - m_B) / (m_A + m_B) \ne 0$). By contrast, very dissymmetric systems such as a neutron-star and a white-dwarf (or a neutron-star and a black hole) will be very efficient emitters of dipolar radiation, and will potentially lead to very strong constraints on tensor-scalar
theories. See below.

\subsection{Theory-space analyses of binary pulsar data}\label{ssec:5.6}

Having reviewed the theoretical results needed to discuss the predictions of alternative gravity theories, let us end by summarizing the results of various theory-space analyses of binary pulsar data.

\smallskip

Let us first recall what are the best, current solar-system limits on the two 1PN `post-Einstein'
parameters $\bar\gamma \equiv \gamma^{\rm PPN} - 1$ and
$\bar\beta \equiv \beta^{\rm PPN} - 1$. They are: 
\begin{equation}
\label{eq4.1:5.45}
\bar\gamma = (2.1 \pm 2.3) \times 10^{-5} \, ,
\end{equation}
from frequency shift measurements made with the Cassini spacecraft \cite{Bertotti}, which supersedes the constraint
\begin{equation}
\label{eq4.2:5.46}
\bar\gamma = (-1.7 \pm 4.5) \times 10^{-4}
\end{equation}
from VLBI measurements \cite{Shapiro-Davis-Lebach-Gregory},
\begin{equation}
\label{eq4.3:5.47}
\vert 2 \, \bar\gamma - \bar\beta \vert < 3 \times 10^{-3} \, ,
\end{equation}
from Mercury's perihelion shift
\cite{Will01,Shapiro-in-Ashby}, and 
\begin{equation}
\label{eq4.4:5.48}
4 \, \bar\beta - \bar\gamma = (4.4 \pm 4.5) \times 10^{-4} \, ,
\end{equation}
from Lunar laser ranging measurements \cite{Williams04}.

\smallskip

Concerning binary pulsar data, we can make use of the published measurements of various Keplerian and post-Keplerian timing parameters in the binary pulsars: PSR 1913$+$16 \cite{WeisbergTaylor04}, PSR B1534$+$12 \cite{Stairs02}, PSR J1141$-$6545 \cite{Bailes03} and PSR J0737$-$3039A$+$B \cite{Lyne04,Kramer04,Krameretal06}. In addition, we can use\footnote{There is, however, a caveat in the theoretical use one can make of the phenomenological limits on $\Delta$. Indeed, in the small-eccentricity long-orbital-period binary pulsar systems used to constrain $\Delta$ one does not have access to enough PK parameters to measure the pulsar mass $m_A$ directly. As the theoretical expression of $\Delta \simeq \alpha_0 \, \alpha_A - \alpha_0^2$ depends on $m_A$ (through $\alpha_A$), one needs to assume some fiducial value of $m_A$ (say $m_A \simeq 1.35 \, M_{\odot}$).} the recently updated limit on the parameter $\Delta$ measuring a possible violation of the strong equivalence principle (SEP), namely $\vert \Delta \vert < 5.5 \times 10^{-3}$ at the 95\% confidence level \cite{Stairsetal}. 

\smallskip

This ensemble of solar-system and binary-pulsar data can then be analyzed within any given parametrized theoretical framework. For instance, one might work within
\begin{enumerate}
\item[(i)] the 4-parameter framework $T_0 (\bar\gamma , \bar\beta ; \epsilon , \zeta)$ \cite{DEF2PN} which defines the 2PN extension of the original (Eddington) PPN framework $T_0 (\bar\gamma , \bar\beta)$; or
\item[(ii)] the 2-parameter class of tensor-mono-scalar theories $T_1 (\alpha_0 , \beta_0)$ \cite{DEF93}; or
\item[(iii)] the 2-parameter class of tensor-bi-scalar theories $T_2 (\beta' , \beta'')$ \cite{DEF92}.
\end{enumerate}

Here, the index $0$ on $T_0 (\bar\gamma , \bar\beta ; \epsilon , \zeta)$ is a reminder of the fact that this framework is not a family of specific theories (it contains {\it zero} explicit dynamical fields), but is a parametrization of 2PN deviations from GR. As a consequence, its use for analyzing binary pulsar data is somewhat ill-defined because one needs to truncate the various timing observables (which are functions of the compactness of the two bodies $A$ and $B$, say $P^{\rm PK} = f(c_A , c_B)$) at the 2PN order (i.e. essentially at the quadratic order in $c_A$ and/or $c_B$). For some observables (or for product of observables) there might be several ways of defining this truncation. In spite of this slight inconvenience, the use of the $T_0 (\bar\gamma , \bar\beta ; \epsilon , \zeta)$ framework is conceptually useful because it shows very clearly why and how binary-pulsar data can probe the behaviour of gravitational theories beyond the usual 1PN regime probed by solar-system tests. 

\smallskip

For instance, the parameter $\Delta_A \equiv m_A^{\rm grav} / m_A^{\rm inert} - 1$ measuring the strong equivalence principle (SEP) violation in a neutron star has, within the $T_0 (\bar\gamma , \bar\beta ; \epsilon , \zeta)$ framework, a 2PN-order expansion of the form \cite{DEF92,DEF2PN}
\begin{equation}
\label{eq5.2n}
\Delta_A = -\frac{1}{2} \, (4 \, \bar\beta - \bar\gamma) \, c_A + \left( \frac{\epsilon}{2} + \zeta + {\mathcal O} (\bar\beta) \right) \, b_A \, ,
\end{equation}
where $c_A = - 2 \, \frac{\partial \ln m_A}{\partial \ln G} \simeq \frac{1}{c^2} \, \langle U \rangle_A$, $b_A = \frac{1}{c^4} \, \langle U^2 \rangle_A \simeq B \, c_A^2$, with $B \simeq 1.026$ and $c_A \simeq k \, m_A / M_{\odot}$ with $k \sim 0.21$. The general result (\ref{eq5.2n}) is compatible with the result quoted in subsection~\ref{ssec:5.2} within the context of the theory $T_2 (\beta' , \beta'')$ when taking into account the fact that, within $T_2 (\beta' , \beta'')$, one has $\bar\beta = \bar\gamma = 0$, $\epsilon = \beta'$ and $\zeta = 0$ [and that $\beta''$ parametrizes some effects beyond the 2PN level]. 

\smallskip

On the example of Eq.~(\ref{eq5.2n}) one sees that, after having used solar-system tests to constrain the first contribution on the RHS to a very small value, one can use binary-pulsar tests of the SEP to set a significant limit on the combination $\frac{1}{2} \, \epsilon + \zeta$ of 2PN parameters. Other pulsar data then yield significant limits on other combinations of the two 2PN parameters $\epsilon$ and $\zeta$. The final conclusion is that binary-pulsar data allow one to set significant limits (around or better than the 1\% level) on the possible 2PN deviations from GR (in contrast to solar-system tests which are unable to yield any limit on $\epsilon$ and $\zeta$) \cite{DEF2PN}. For a recent update of the limits on $\epsilon$ and $\zeta$, which makes use of recent pulsar data see \cite{DEF06}.

\smallskip

Let us now briefly discuss the use of mini-space of theories, such as $T_1 (\alpha_0 , \beta_0)$ or $T_2 (\beta' , \beta'')$, for analyzing solar-system and binary-pulsar data. The basic methodology is to compute, for each given theory (e.g. for each given values of $\alpha_0$ and $\beta_0$ if one chooses to work in the $T_1 (\alpha_0 , \beta_0)$ theory space) a goodness-of-fit statistics $\chi^2 (\alpha_0 , \beta_0)$ measuring the quality of the agreement between the experimental data and the considered theory. For instance, when considering the timing data of a particular pulsar, for which one has measured several PK parameters $p_i$ ($i=1,\ldots , n$) with some standard deviations $\sigma_{p_i}^{\rm obs}$, one defines, for this pulsar
\begin{equation}
\label{eq3.29}
\chi^2 (\alpha_0 , \beta_0) = \underset{m_A , m_B}{\min}  \sum_{i=1}^n (\sigma_{p_i}^{\rm obs})^{-2} (p_i^{\rm theory} (\alpha_0 , \beta_0 ; m_A , m_B) - p_i^{\rm obs})^2 \, ,
\end{equation}
where `min' denotes the result of minimizing over the unknown masses $m_A , m_B$ and where $p_i^{\rm theory} (\alpha_0 , \beta_0 ; m_A , m_B)$ denotes the theoretical prediction (within $T_1 (\alpha_0 , \beta_0)$) for the PK observable $p_i$ (given also the observed values of the Keplerian parameters).

\smallskip

The goodness-of-fit quantity $\chi^2 (\alpha_0 , \beta_0)$ will reach its minimum $\chi_{\rm min}^2$ for some values, say $\alpha_0^{\rm min} , \beta_0^{\rm min}$, of  $\alpha_0$ and $\beta_0$. Then, one focusses, for each pulsar, on the level contours of the function
\begin{equation}
\label{eq3.32}
\Delta \, \chi^2 (\alpha_0 , \beta_0) \equiv \chi^2 (\alpha_0 , \beta_0) - \chi_{\rm min}^2 \, .
\end{equation}
Each choice of level contour (e.g. $\Delta \, \chi^2 = 1$ or $\Delta \, \chi^2 = 2.3$) defines a certain region in theory space, which contains, with a certain corresponding `confidence level', the `correct' theory of gravity (if it belongs to the considered mini-space of theories). When combining together several independent data sets (e.g. solar-system data, and different pulsar data) we can define a total goodness-of-fit statistics $\chi_{\rm tot}^2 (\alpha_0 , \beta_0)$, by adding together the various individual $\chi^2 (\alpha_0 , \beta_0)$. This leads to a corresponding combined contour $\Delta \, \chi^2_{\rm tot} (\alpha_0 , \beta_0)$.

\smallskip

Let us end by briefly summarizing the results of the theory-space approach to relativistic gravity tests. For detailed discussions the reader should consult Refs.~\cite{DEF92,TWDW92,DEF96,DEF98,Esposito-Farese05}, and especially the recent update \cite{DEF06} which uses the latest binary-pulsar data.

\smallskip

Regarding the two-parameter class of tensor-bi-scalar theories $T_2 (\beta' , \beta'')$ the recent analysis \cite{DEF06} has shown that the $\Delta \, \chi^2 (\beta' , \beta'')$ corresponding to the double binary pulsar PSR J0737$-$3039 was defining quite a small elliptical allowed region in the $(\beta',\beta'')$ plane. By contrast the other pulsar data define much wider allowed regions, while the strong equivalence principle tests define (in view of the theoretical result $\Delta \simeq 1 + \frac{1}{2} \, B\beta' (c_A^2 - c_B^2)$) a thin, but infinitely long, strip $\vert \beta' \vert < {\rm cst.}$ in the $(\beta',\beta'')$ plane. This highlights the power of the double binary pulsar in probing certain specific strong-field deviations from GR.

\smallskip

\begin{figure}[h]
\centering
\includegraphics[height=65mm]{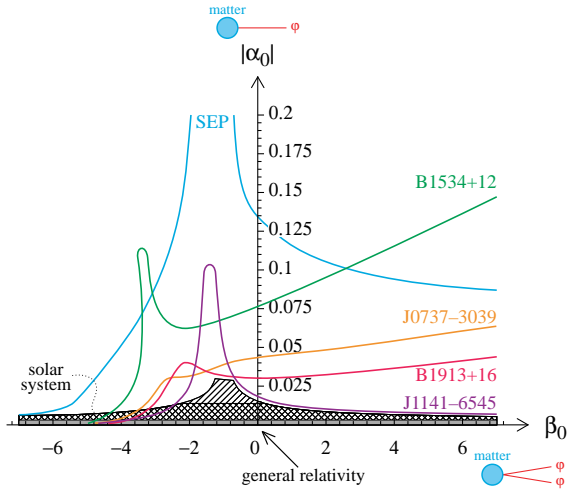}
%
%
\caption{Solar-system and binary-pulsar constraints on the two-parameter family of tensor-mono-scalar theories $T_1 (\alpha_0 , \beta_0)$. Figure taken from \cite{DEF06}.}
\label{fig:3}       
\end{figure}

Contrary to the $T_2 (\beta' , \beta'')$ tensor-bi-scalar theories, which were constructed to have exactly the same first post-Newtonian limit as GR\footnote{However, this could be achieved only at the cost of allowing some combination of the two scalar fields to carry a negative energy flux.} (so that solar-system tests put no constraints on $\beta'$ and $\beta''$), the class of tensor-mono-scalar theories $T_1 (\alpha_0 , \beta_0)$ is such that its parameters $\alpha_0$ and $\beta_0$ parametrize {\it both} the weak-field 1PN regime (see Eqs.~(\ref{eq2.19a:5.20a}) and (\ref{eq2.19b:5.20b}) above) and the strong-field regime (which plays an important role in compact binaries). This means that each class of solar-system data (see Eqs.~(\ref{eq4.1:5.45})--(\ref{eq4.4:5.48}) above) will define, via a corresponding goodness-of-fit statistics of the type, say
$$
\chi_{\rm Cassini}^2 (\alpha_0 , \beta_0) = (\sigma_{\gamma}^{\rm Cassini})^{-2} \, (\bar\gamma^{\rm theory} (\alpha_0 , \beta_0) - \bar\gamma^{\rm Cassini})^2
$$
a certain allowed region\footnote{Actually, in the case of the Cassini data, as it is quite plausible that the {\it positive} value of the published central value $\bar\gamma^{\rm Cassini} = + 2.1 \times 10^{-5}$ is due to unsubtracted systematic effects, we use $\sigma_{\gamma}^{\rm Cassini} = 2.3 \times 10^{-5}$ but $\bar\gamma^{\rm Cassini} = 0$. Otherwise, we would get unreasonably strong $1\sigma$ limits on $\alpha_0^2$ because tensor-scalar theories predict that $\bar\gamma$ must be {\it negative}, see Eqs.~(\ref{eq2.19a:5.20a}) and (\ref{eq2.19b:5.20b}).} in the $(\alpha_0 , \beta_0)$ plane. As a consequence, the analysis in the framework of the $T_1 (\alpha_0 , \beta_0)$ space of theories allows one to compare and contrast the probing powers of solar-system tests versus binary-pulsar tests (while comparing also solar-system tests among themselves and binary-pulsar ones among themselves). The result of the recent analysis \cite{DEF06} is shown in Figure~\ref{fig:3}.

%
%

In Fig.~\ref{fig:3}, the various solar-system constraints (\ref{eq4.1:5.45})--(\ref{eq4.4:5.48}) are concentrated around the horizontal $\beta_0$ axis. In particular, the high-precision Cassini constraint is the lower small grey strip. The various pulsar constraints are labelled by the name of the pulsar, except for the strong equivalence principle constraint which is labelled SEP. Note that General Relativity corresponds to the origin of the $(\alpha_0 , \beta_0)$ plane, and is compatible with all existing tests.

\smallskip

The global constraint obtained by combining all the pulsar tests would, to a good accuracy, be obtained by intersecting the various pulsar-allowed regions. One can then see on Fig.~\ref{fig:3} that it would be comparable to the pre-Cassini solar-system constraints and that its boundaries would be defined successively (starting from the left) by 1913$+$16, 1141$-$6545, 0737$-$3039, 1913$+$16 again and 1141$-$6545 again. 

\smallskip

A first conclusion is therefore that, at the {\it quantitative level}, binary-pulsar tests constrain tensor-scalar gravity theories as strongly as most solar-system tests (excluding the exceptionally accurate Cassini result which constrains $\alpha_0^2$ to be smaller than $1.15 \times 10^{-5}$, i.e. $\vert \alpha_0 \vert < 3.4 \times 10^{-3}$). A second conclusion is obtained by comparing the behaviour of the solar-system exclusion plots and of the binary-pulsar ones around the negative $\beta_0$ axis. One sees that binary-pulsar tests exclude a whole domain of the theory space (located on the left of $\beta_0 < -4$) which is compatible with all solar-system experiments (even when including the very tight Cassini constraint). This remarkable {\it qualitative} feature of pulsar tests is a direct consequence of the existence of (non-perturbative) {\it strong-field effects} which start developing when the product $-\beta_0 \, c_A$ (with $c_A$ denoting, as above, the compactness of the pulsar) becomes of order unity.

\section{Conclusion}\label{sec:6}

In conclusion, we hope to have convinced the reader of the superb opportunities that binary pulsar data offer for testing gravity theories. In particular, they have been able to go qualitatively beyond solar-system experiments in probing two physically important regimes of relativistic gravity: the radiative regime and the strong-field one. Up to now, General Relativity has passed with flying colours all the radiative and strong-field tests provided by pulsar data. However, it is important to continue testing General Relativity in all its aspects (weak-field, radiative and strong-field). Indeed, history has taught us that physical theories have a limited range of validity, and that it is quite difficult to predict in which regime a theory will cease to be an accurate description of nature. Let us look forward to new results, and possibly interesting surprises, from binary pulsar data.

\bigskip

\noindent {\bf Acknowledgments}

\smallskip

\noindent It is a pleasure to thank my long-term collaborator Gilles Esposito-Far\`ese for his useful remarks on the text, and for providing the figures. I wish also to thank the organizers of the 2005 Sigrav School, and notably Monica Colpi and Ugo Moschella, for organizing a warm and intellectually stimulating meeting. This work was partly supported by the European Research and Training Network ``Forces Universe'' (contract number MRTN-CT-2004-005104).

%



\end{document}